\documentclass{emulateapj}

\def\kmsec{\mbox{km~s$^{\rm -1}$}}

\def\rpro{\mbox{$r$-process}}
\def\spro{\mbox{$s$-process}}
\def\ncap{\mbox{$n$-capture}}

\shorttitle{Stellar Composition of Milky Way Halos}
\shortauthors{I.~U.\ Roederer}

\begin{document}

\title{Chemical Inhomogeneities in the Milky Way Stellar Halo}

\author{Ian U.\ Roederer}
\affil{Department of Astronomy, University of Texas at Austin \\
1 University Station C1400, Austin, TX 78712-0259}
\email{iur@astro.as.utexas.edu}

\begin{abstract}

We have compiled a sample of 699 stars from the recent literature
with detailed chemical abundance information
(spanning $-4.2 \lesssim$~[Fe/H]~$\lesssim +0.3$), and we compute their
space velocities and Galactic orbital parameters.
We identify members of the inner and outer stellar halo populations 
in our sample based only on their kinematic properties 
and then compare the abundance ratios of these populations
as a function of [Fe/H].
In the metallicity range where the two populations overlap
($-2.5 \lesssim$~[Fe/H]~$\lesssim -1.5$),
the mean [Mg/Fe] of the outer halo is lower than 
the inner halo by $\sim$~0.1~dex.
For [Ni/Fe] and [Ba/Fe], 
the star-to-star abundance scatter of the inner halo is consistently
smaller than in the outer halo.
The [Na/Fe], [Y/Fe], [Ca/Fe], and [Ti/Fe] ratios of both populations 
show similar means and levels of scatter.
Our inner halo population is chemically homogeneous, 
suggesting that a significant fraction
of the Milky Way stellar halo originated from a well-mixed ISM.
In contrast, our outer halo population is chemically diverse,
suggesting that another significant fraction
of the Milky Way stellar halo formed in remote regions where
chemical enrichment was dominated by local supernova events.
We find no abundance trends with maximum radial distance from
the Galactic center or maximum vertical distance from the Galactic disk.
We also find no common kinematic signature for groups of
metal-poor stars with peculiar abundance patters, such as 
the $\alpha$-poor stars or 
stars showing unique neutron-capture enrichment patterns.
Several stars and dSph systems with unique abundance patterns 
spend the majority of their time in the distant
regions of the Milky Way stellar halo, suggesting that the 
true outer halo of the Galaxy may have little resemblance
to the local stellar halo.

\end{abstract}


\section{Introduction}
\label{introduction}

The nucleosynthesis reactions necessary to 
produce metals in stars were realized nearly half a century ago
\citep{burbidge57,seeger65,fowler67,wagoner67}, yet the challenge to
interpret the wide variety of nucleosynthetic signatures 
observed in different stellar populations today remains as
intriguing as ever.
Metal-poor stellar populations should contain
recycled stellar material from fewer generations of stars 
than metal-rich populations, making interpretation of their
chemical enrichment history---\textit{in principle}---simpler.
Thanks to numerous large surveys over the last four decades
designed to identify metal-poor stars (see review by \citealt{beers05}),
the tally of known metal-poor stars now stretches well into the thousands.
Concurrently, great advances have been made in the analysis and
interpretation of the chemical signatures and 
enrichment histories revealed by stellar spectra 
(e.g., \citealt{audouze76}, 
\citealt{kraft79}, \citealt{wheeler89}, 
\citealt{mcwilliam97}, \citealt{gratton04}, 
\citealt{beers05}, and \citealt{sneden08}).
Furthermore, careful laboratory analysis has improved 
our knowledge of the relevant atomic data necessary to make
accurate and detailed records of the chemical composition of the
atmospheres of metal-poor stars.
The confluence of advances in each of these fields
has built upon the foundation of stellar nucleosynthesis
to greatly increase our
understanding of the earliest generations of stars, 
the chemical evolution of the various components of the Milky Way Galaxy, 
and the formation process of the Galaxy, to name just a few successes.

Correlations between metal enrichment in stellar populations 
and the kinematic properties of these stars have been known for
some time now \citep[e.g.,][]{eggen62,wallerstein62}, 
and through the years these relationships 
have been fleshed out in increasing detail.
Most studies of the formation of the stellar halo of the Galaxy 
have employed limited chemical data (e.g., [Fe/H]\footnote{
We adopt the usual spectroscopic notations that
[A/B]~$\equiv$ log$_{10}$(N$_{\rm A}$/N$_{\rm B}$)$_{\star}$~--
log$_{10}$(N$_{\rm A}$/N$_{\rm B}$)$_{\odot}$ and
log~$\epsilon$(A)~$\equiv$ 
log$_{10}$(N$_{\rm A}$/N$_{\rm H}$)~$+$~12.00
for elements A and B.}) to accompany the kinematic data.
This is sufficient to study the metallicity distribution
function (MDF) of the halo 
(e.g., \citealt{hartwick76,ryan91b,ivezic08,schorck08}),
formation and age of the halo 
(e.g., \citealt{eggen62,searle78,sandage86,wyse88,
gilmore89,preston91,ryan91a,majewski92,norris94,carney96,
sommerlarsen97,chiba98,chiba00,carollo07,bell08,miceli08,morrison08}),
or for investigating stellar streams and halo substructure
(e.g., \citealt{majewski96,helmi99,chiba00,gilmore02,kinman07}).
Iron (Fe)---or, in some cases, calcium (Ca)---alone 
is less useful for studies of the 
chemical enrichment of the halo, which can examine, e.g., 
supernova (SN) models and rates, 
stellar binary fractions, 
mixing processes in the interstellar medium (ISM) of the halo,
the Galactic potential, Galactic structure, 
and relationships between various substructures
(e.g., globular clusters, Local Group dwarf spheroidal [dSph]
systems, stellar streams).
It was not until recently that kinematic and detailed chemical data 
for halo stars were analyzed together
\citep[e.g.,][]{gratton03,simmerer04,venn04,pritzl05,font06,geisler07}.

Three substantial advances have been made in the short period of time since 
\citet{venn04} and \citet{pritzl05} performed
a detailed chemical comparison between field stars in the halo and
nearby dSph systems and globular clusters.
First, echelle spectrographs on multiple 
large (6--10~m class) telescopes have enabled
investigators to carry out detailed abundance analyses of large numbers
of (often faint) metal poor stars, including many stars with
[Fe/H]~$<-3.0$.
Second, investigators at the US Naval Observatory 
have released revised proper motion catalogs based on
longer time baselines, improved quality of astrographs, 
and better techniques for digitizing 
photographic plates \citep[e.g.,][]{zacharias04a}.
Improved proper motions are given in these catalogs for many of the
metal-poor stars investigated in recent years, 
enabling us to derive their full space motions through the Galaxy.
Finally, large numbers ($\sim$~few~$\times$~10$^{4}$) of calibration
stars with known stellar parameters and metallicities have been
observed as part of the Sloan Digital Sky Survey (SDSS) and 
Sloan Extension for Galactic Understanding and Exploration (SEGUE)
projects \citep[e.g.,][]{newberg03,allendeprieto07,lee07}.
This has enabled the previously known kinematic properties
of nearby members of the so-called inner and outer stellar halos 
to be assessed with a new level of detail \citep{carollo07}.
These advances insist on a fresh reanalysis of the existing data.
The goal of the present study is to interpret the wealth of
recent high-resolution abundance analyses of metal-poor stars
in light of the most recent kinematic knowledge of these two
major components of the Galactic halo.

\section{Abundance Data from the Literature}
\label{litdata}

\citet{venn04} compiled from existing literature 
a large sample of stellar abundances and
\textit{UVW} kinematic data, when available, to compare the 
chemical enrichment patterns of metal-poor Galactic halo stars with
present-day dSph systems.  
We adopt their data from the high-resolution abundance analyses of
\citet{edvardsson93},
\citet{nissen97}, 
\citet{hanson98}, 
\citet{prochaska00}, 
\citet{fulbright00,fulbright02},
\citet{stephens02}, 
\citet{bensby03}, and 
\citet{reddy03}.
The Milky Way thin and thick disks are represented in this sample
along with the halo, totaling 620 stars.

We supplement this sample with abundances derived from more
recent high-resolution analyses of metal-poor halo stars 
or older studies that did not include any kinematic analysis.
We add another 309 stars from the analyses of 
\citet{mcwilliam95a,mcwilliam95b},
\citet{hill02},
\citet{ivans03},
\citet{cayrel04},
\citet{honda04a,honda04b},
\citet{barklem05},
\citet{honda06},
\citet{francois07}, and
\citet{lai08}.
For stars without sufficient kinematic information to compute 
\textit{UVW} velocities and Galactic orbital parameters
(i.e., only the stellar radial velocity is published), we obtain 
proper motions from the catalogs listed in \S~\ref{uvw}.
Many of these stars are not in the
Hipparcos Catalog \citep{perryman97}, and for these stars we derive a 
photometric parallax (see \S~\ref{distances}).
We also require that the total proper motion 
is greater than 2.5 times its error, 
otherwise we discard the star from further kinematic analysis.
We insert this requirement to derive accurate
space velocities, yet unfortunately this will also bias our 
sample against stars with small proper motions;
see \S~\ref{caution} for closer analysis of this point.
This requirement culls our sample by 230 stars. 
Our final sample of stars is given in Table~\ref{startab}
along with their adopted [Fe/H] values.

This large dataset will allow us to probe abundance trends as a function
of stellar kinematics.
Because we mix data from different sources,
systematic offsets in the abundance ratios arise 
from to a variety of factors, including
(1) different spectral resolution and signal-to-noise (S/N) ratios 
of the data themselves, leading to potential blending of lines
in crowded spectral regions, 
since high resolution is desirable to more-fully resolve the absorption
line profiles yet such high resolution and high S/N 
is impractical for studies of very faint stars;
(2) different sets of absorption lines and transition probabilities
used in each study;
(3) different methods employed to determine stellar parameters;
(4) differences in the structure of the model atmospheres themselves;
and so on.
These sources of systematic error should
be no larger than 0.1--0.2~dex in [X/Fe],
which does limit our ability to detect subtle chemical signatures; 
gross trends should still be identified reliably.
At any given [Fe/H], our sample is comprised of stars from a variety
of studies, and the characteristic scatters of the inner and outer 
halo populations are distinct beyond this level of systematic scatter.

\section{Kinematic Data}
\label{kinematics}

\subsection{Distance Estimates}
\label{distances}

Hipparcos \citep{perryman97} geometric parallaxes are the preferred method
for deriving accurate distances for our sample (using the 
new reduction of the Hipparcos data described in \citealt{vanleeuwen07}).
For stars not in the Hipparcos catalog, we derive distance estimates
from the photometric parallax
by comparing the observed $V$ magnitude with a predicted absolute
magnitude $M_{V}$, derived from the dereddened $V-K$ color.\footnote{
This method was only used to derive distances to stars with
[Fe/H]~$< -1.0$.}
To establish the relationship between color and absolute magnitude, 
we use the $Y^{2}$ set of theoretical isochrones \citep{demarque04}.
We select the appropriate isochrone for each star, 
assuming [$\alpha$/Fe]~$=+0.3$ for all stars and 
choosing a metallicity close to the individual stellar [Fe/H].
We find that the choice of (old) age has negligible effect on our
derived photometric distances (comparing between 10, 11, 12, and 13~Gyr
isochrones), so we adopt an age of 12~Gyr for all stars.
To break the degeneracy of $M_{V}$ with $V-K$ (since any given isochrone
is not single-valued in $M_{V}$ for each value of $V-K$),
we establish the evolutionary state of each star based on the
surface gravity estimate given in the literature
(``dwarfs'': $\log g \geq 4.0$; ``giants'': $\log g < 4.0$).
We adopt $V$ magnitudes from SIMBAD if not given in the references
in \S~\ref{litdata}, $K$ magnitudes from 2MASS\footnote{
This publication makes use of data products from the Two Micron 
All Sky Survey, which is a joint project of the University of 
Massachusetts and the Infrared Processing and Analysis 
Center/California Institute of Technology, funded by the 
National Aeronautics and Space Administration and the 
National Science Foundation.} \citep{cutri03},
reddening estimates from the dust maps of \citet{schlegel98}
and the reddening laws of \citet{mccall04}:
$R_{V} = 3.070 \times E(B-V)$ and $R_{V-K} = 2.727 \times E(B-V)$.
We use the prescription given by \citet{bonifacio00b} to reduce
the \citet{schlegel98} estimates for $E(B-V) > 0.10$:
$E(B-V)_{\rm revised} = 0.10 + 0.65[E(B-V)_{\rm S98} - 0.10]$.

\begin{figure}
\epsscale{1.15}
\plotone{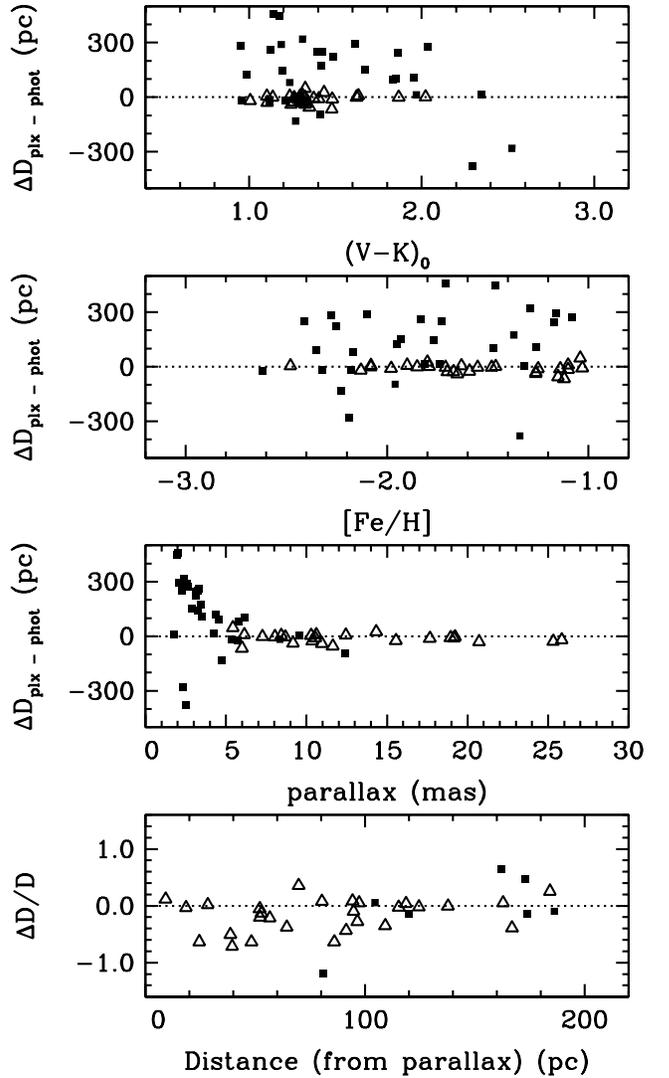}
\caption{
\label{distanceplot}
Comparison of distances derived from Hipparcos parallax 
measurements and photometric parallaxes, as a function of
($V-K$)$_{0}$, [Fe/H], and Hipparcos parallax.
The bottom panel shows the relative differences as a function
of distance (as derived from the Hipparcos parallax) for 
stars with $\pi_{\rm Hipparcos} > 5$~mas~yr$^{-1}$.
Open triangles represent our dwarf sample ($\log g \geq 4.0$) 
and filled squares represent our giant sample ($\log g < 4.0$).
}
\end{figure}

Figure~\ref{distanceplot} compares our photometric and geometric distances
for stars in \citet{venn04}.
In order to assess the reliability of our photometric distances alone,
we attempt to minimize the effect of other influences on the result
by restricting this comparison to stars with
$E(B-V) < 0.30$ and [Fe/H]~$< -1.0$.
The difference in the distance computed by the two methods is shown 
as a function of $V-K$ color, [Fe/H] metallicity,
and Hipparcos parallax.
While it is apparent that there
is poor agreement for the distance estimates for our giants, there exists
no trend between the differences and color or metallicity.
There is, however, a very clear correlation 
between these differences and the Hipparcos parallax.
In other words, 
the agreement is very good for stars with reliable Hipparcos
parallax measurements, but this correlation breaks down for stars with 
$\pi_{\rm Hipparcos} \lesssim 5$~mas~yr$^{-1}$. 
(cf.\ Figure~1 of \citealt{chiba98}, 
\S~3.3.5 and Figure~3 of \citealt{beers00}).
This preferentially affects distant stars (giants, in our sample)
that have parallax measurement uncertainties comparable to the
size of the parallaxes themselves.
The vast majority of stars with $\pi_{\rm Hipparcos} > 5$~mas~yr$^{-1}$ 
are dwarfs, but the few giants with large parallaxes exhibit generally
good agreement between the two distance estimates.
We also show the relative differences as a function of distance
in the bottom panel of Figure~\ref{distanceplot}.
The scatter in the giants ($\sigma = 0.59$, 7 stars)
is about twice as large as the scatter in the dwarfs 
($\sigma = 0.29$, 28 stars), which is to be expected since
a given uncertainty in $(V-K)_{0}$ will translate to a 
relatively larger uncertainty in $M_{V}$ for stars on the 
giant branch than for stars on the main sequence or near the turn-off.
As we proceed, we calculate geometric distances 
if the Hipparcos parallax is greater than 5 times its uncertainty, 
otherwise we estimate distance from the photometric parallax.
These distances are reported in Table~\ref{startab}.

\subsection{Space Velocities}
\label{uvw}

We calculate $UVW$ space velocities for the stars in our sample 
from measurements of the radial velocity, proper motion, and distance
to each star.
We adopt a left-handed coordinate system, where
positive $U$ denotes velocity towards the Galactic anti-center
($\ell = 180^{\circ}$, $b = 0^{\circ}$),
positive $V$ denotes velocity parallel to the direction of rotation
of the Local Standard of Rest (LSR; $\ell = 90^{\circ}$, $b = 0^{\circ}$),
and positive $W$ denotes velocity perpendicular to the plane of the
disk of the galaxy ($b = +90^{\circ}$).
We correct for the motion of the Sun with respect to the LSR, adopting
($U_{\rm LSR}$, $V_{\rm LSR}$, $W_{\rm LSR}$)$_{\odot}=$~($-9$, $12$, $7$) 
\citep{mihalas81}, 
where all velocities are measured in km\,s$^{-1}$.
We adopt $V_{\rm LSR} = 220$\,km\,s$^{-1}$ \citep{kerr86} and 
report the $V$ component in a Galactic reference frame (i.e., 
where the Sun's motion is described by [$U$, $V$, $W$] $=$
[$-9$, $232$, $7$]).
Radial velocities were taken from the literature references
described in \S~\ref{litdata}.
Proper motions were taken from the second data
release of the USNO CCD Astrograph Catalog (UCAC2; 
\citealt{zacharias04a,urban04})
or the USNO-B astrometry catalog\footnote{
Accessible through the Naval Observatory Merged Astrometric Dataset,
NOMAD \citep{zacharias04b}.} \citep{monet03},
with preference given to the UCAC2 measurements.

The $UVW$ velocities, measured in a Cartesian coordinate system,
can be transformed into a cylindrical coordinate system 
($R$, $\phi$, $z$) with its origin at the Galactic center, 
where $R = \sqrt{(R_{\odot}+x)^{2}+y^{2}}$, 
$\phi = \arctan(y/(R_{\odot}+x))$, and $z=z$.
Our adopted and derived $UVW$ and $V_{\phi}$ velocities are 
given in Table~\ref{startab}.

\subsection{Galactic Orbit Parameters}
\label{orbit}

We use the orbit integrator developed by D.~Lin to calculate 
the Galactic orbital parameters for stars in our sample.
The gravitational potential of the Galaxy is computed by summing
the individual contributions of the disk, spheroid (or bulge), 
and halo components:
\begin{mathletters}
\begin{eqnarray}
\Phi_{\rm disk} & = & \frac{-GM_{\rm disk}}
  {\sqrt{R^{2} + (a + \sqrt{z^{2} + b^{2}})^{2}}} \\
\Phi_{\rm spheroid} & = & \frac{-GM_{\rm spheroid}}{r+c} \\
\Phi_{\rm halo} & = & v^{2}_{\rm halo} \ln{(r^{2} + d^{2})}, 
\end{eqnarray}
\label{ourpotential}
\end{mathletters}
where 
$R = \sqrt{x^{2} + y^{2}}$ and 
$r = \sqrt{x^{2} + y^{2} + z^{2}}$,
$M_{\rm disk} = 10^{11}$\,M$_{\sun}$,
$M_{\rm spheroid} = 3.4 \times 10^{10}$\,M$_{\sun}$,
$v_{\rm halo} = 128$\,km\,s$^{-1}$,
$a = 6.5$, 
$b = 0.26$, 
$c = 0.7$, and 
$d = 12.0$. 
This model is described in detail in \citet{johnston96} and \citet{johnston98},
who chose values for these parameters 
to reproduce a nearly-flat rotation curve for the Milky Way Galaxy 
between $r = 1$ and $r = 30$\,kpc and a disk scale height of 0.2\,kpc.
We require that each star complete a minimum of 20 orbits around the
Galactic center in 10~Gyr (at which point the orbital parameters
have settled to a constant value); no orbital parameters are given
in Table~\ref{startab} for stars that did not meet this requirement.
From the output we calculate 
$r_{\rm peri}$ and $r_{\rm apo}$, 
the minimum and maximum Galactocentric radii reached by each star; 
$|Z_{\rm max}|$, the maximum distance above (or below) the Galactic plane; 
and $e$, the eccentricity of the orbit, defined as 
$(r_{\rm apo}-r_{\rm peri}) / (r_{\rm apo}+r_{\rm peri})$.
These properties are also reported in Table~\ref{startab}.

\begin{figure*}
\epsscale{1.00}
\plotone{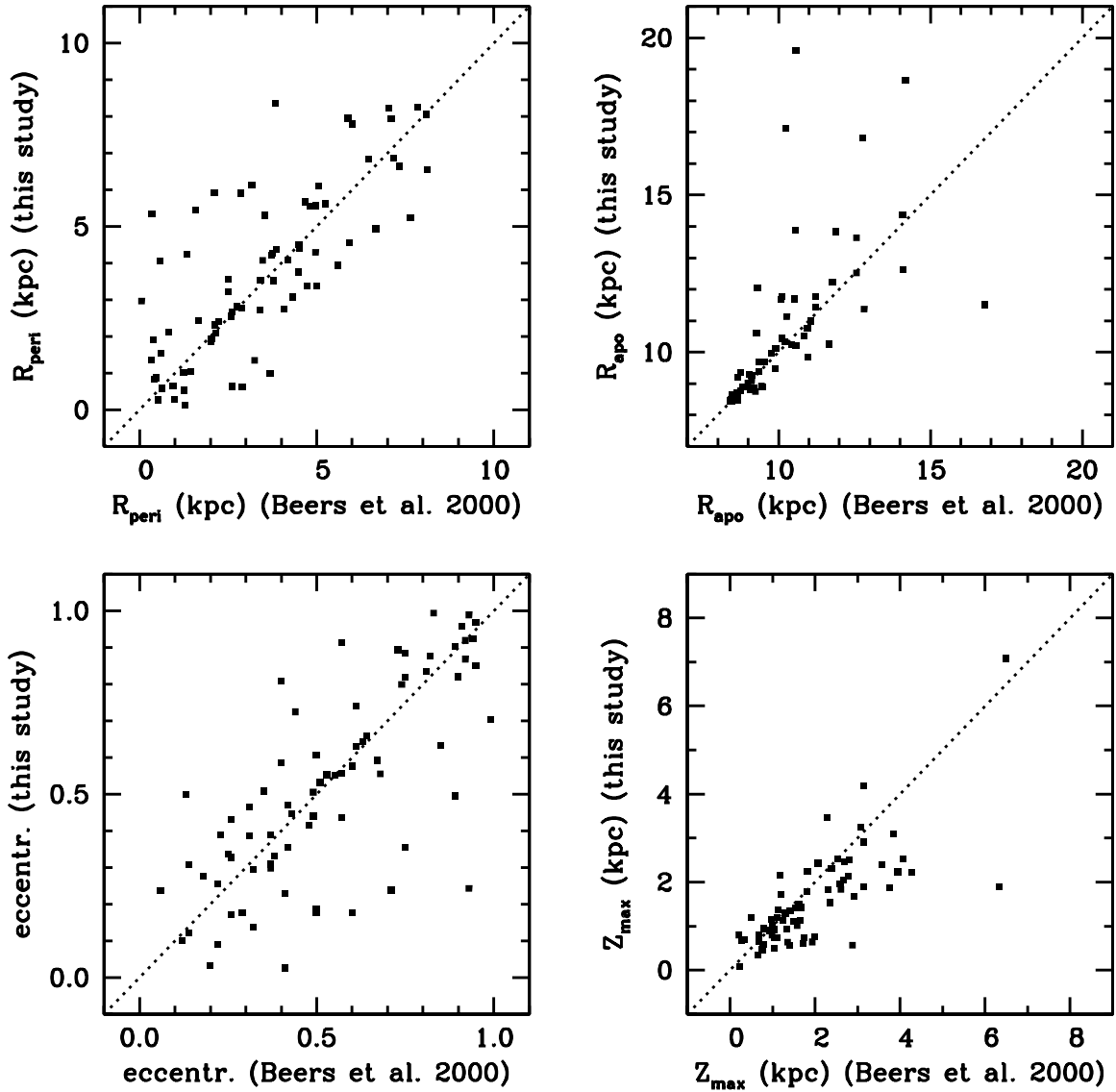}
\caption{
\label{compareb00}
Comparison of our predicted Galactic orbit properties 
with those of \citet{beers00}.
Our values have been computed using the three-component model
for the potential, while \citet{beers00} used the analytic 
St\"{a}ckel-type potential.
}
\end{figure*}

Our kinematic results ought to be reasonably independent of our 
model for the Galactic potential.
Since we adopt the Galactic orbital properties of the inner and outer
halo populations from \citet{carollo07} (see \S~\ref{halokinematics}), 
we want to demonstrate some degree of consistency between 
results derived from the two different potentials.
To assess this, we compare the Galactic orbital properties 
of stars computed using Equation~\ref{ourpotential} 
to those computed using the analytic St\"{a}ckel-type potential 
\citep{sommerlarsen90,chiba98} adopted by \citet{beers00}
and \citet{carollo07}.
The parameters of their model have been tuned to reproduce a 
flat Galactic rotation curve beyond $R=4$~kpc and the local mass 
density at the Solar radius.
We use our three-component model of the Galactic potential to 
compute Galactic orbital parameters for 100 stars from the sample
of \citet{beers00}, strictly using their input data (distance, 
radial velocity, proper motion, etc.).
Figure~\ref{compareb00} compares the orbital parameters presented
in \citet{beers00} (i.e., computed with the St\"{a}ckel-type potential)
with those derived with our three-component model.
Both models generally produce similar apogalactic and perigalactic
distances 
as well as maximum vertical distance, but with a fair degree of scatter
in each of these quantities.
For $R_{\rm apo} > 10$~kpc and $|Z_{\rm max}| > 2$~kpc, the 
distances for individual stars are typically different by less than
a factor of two.
These systematic differences are inherent to the representations of the
potential.

How worrisome are these differences?
Our approach to identifying the two populations isolates them
from one another in kinematic space by placing a buffer between 
them---reasonable uncertainty in the distance estimate or proper motion
of a star should not cause it to drift from one classification to
the other.
For example, stars in our inner halo population must not orbit more than 
15~kpc from the Galactic center, while stars in our outer halo population
must orbit to at least 25~kpc.
We also adopt membership standards for the inner halo that 
rely on more than one kinematic selection criterion.
Therefore, we do not consider this systematic difference a serious problem
in our kinematic selection criteria.

\section{New Kinematic Definitions of the Inner and Outer Halo Populations}
\label{halokinematics}

Analyzing more than 10,000 calibration stars from low resolution 
($R \equiv \lambda/\Delta\lambda \sim 2000$) spectra
obtained with the SDSS, \citet{carollo07} reported that the
Milky Way stellar halo is roughly divisible into two
broadly-overlapping components.
Their ``inner halo'' is composed of stars on highly eccentric orbits,
has a flattened density distribution,
exhibits a modest net prograde rotation, and
has a peak in the MDF at [Fe/H]~$= -1.6$.
Their ``outer halo'' is composed of stars with a variety of orbital
eccentricities, has a spherical density distribution,
exhibits a net retrograde rotation, and has a peak in the
MDF at [Fe/H]~$= -2.2$.
While their inner halo primarily dominates the stellar halo
at Galactocentric radii $\lesssim$~10--15~kpc and the
outer halo primarily dominates at radii $\gtrsim$~15--20~kpc,
these two components overlap in the Solar neighborhood,
disguising their distinct characteristics from being
recognized by earlier, smaller samples of stars.

\begin{figure*}
\epsscale{1.00}
\plotone{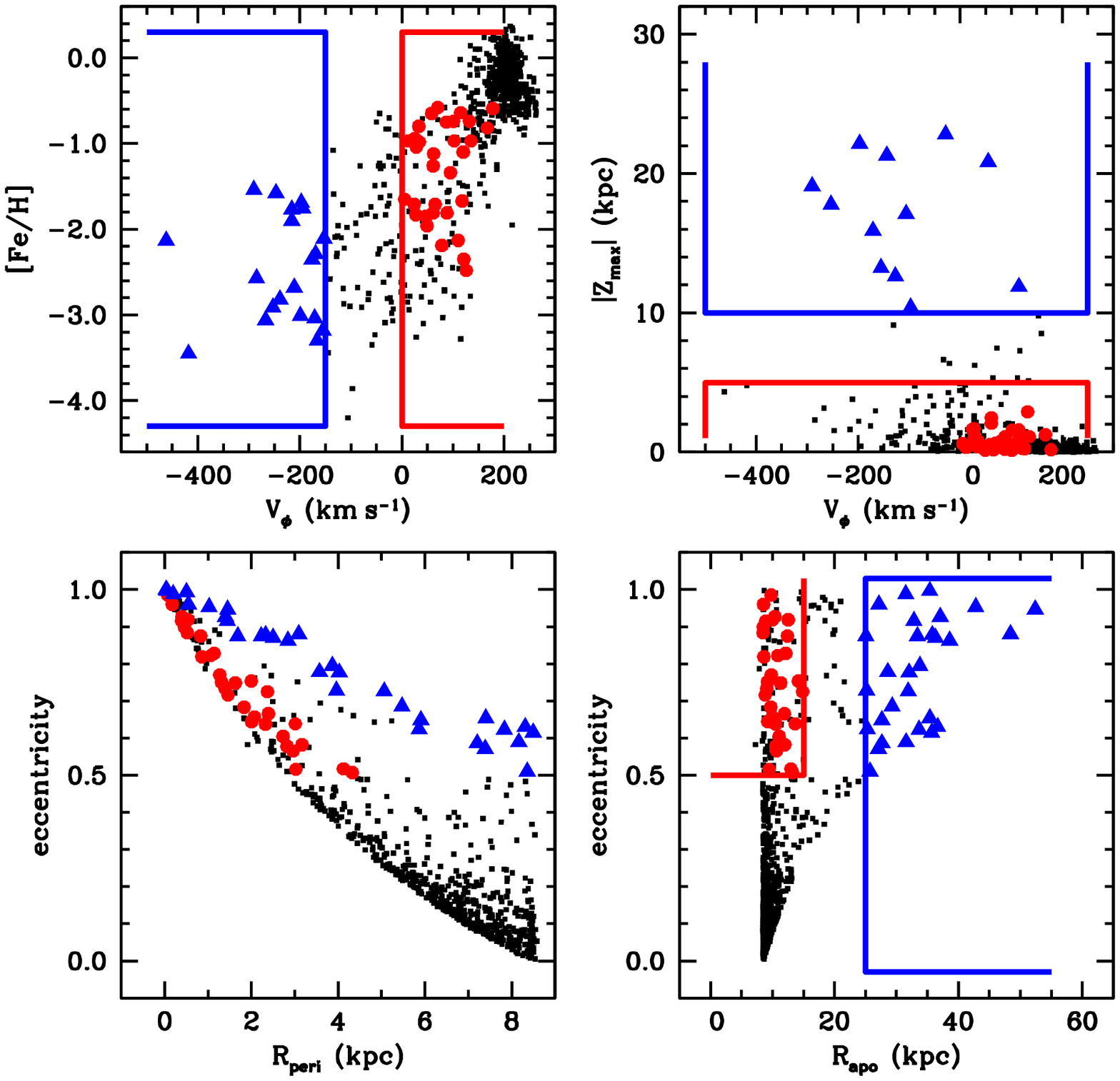}
\caption{
\label{innerouterplot}
The kinematic definitions of our inner and outer halo samples.
The inner halo population (dark gray circles, red in the online edition) 
is defined by stars
that have $V_{\phi} > 0$~\kmsec, $|Z_{\rm max}| < 5$~kpc, 
$R_{\rm apo} < 15$~kpc, and $e > 0.5$.
The outer halo population (black triangles, blue in the online edition) 
is defined by stars
that have $V_{\phi} < -150$~\kmsec, $|Z_{\rm max}| > 10$~kpc,
or $R_{\rm apo} > 25$~kpc.
Stars that did not meet the kinematic criteria for these two 
populations are shown as small dots.
[Please see the electronic edition for a color version
of this figure.]
}
\end{figure*}

We define membership in the inner and outer halo populations 
based \textit{only} on the kinematics of stars in our sample.
These new criteria, based on the classifications sketched by 
\citet{carollo07}, are shown in Figure~\ref{innerouterplot}.
For the remainder of the present study, we shall use the working
definition that 
inner halo membership is characterized by prograde rotation about
the Galactic center ($V_{\phi} > 0$~\kmsec), 
relatively small maximum vertical distance from the Galactic plane
($|Z_{\rm max}| < 5$~kpc), relatively small maximum radial distance
from the Galactic center ($R_{\rm apo} < 15$~kpc), and
rather high orbital eccentricity ($e > 0.5$).
A star must meet each of these criteria to be classified as a 
member of the inner halo.
Outer halo membership is characterized by significant retrograde
rotation about the Galactic center ($V_{\phi} < -150$~\kmsec),
relatively large maximum vertical distance from the Galactic plane
($|Z_{\rm max}| > 10$~kpc), or relatively large maximum radial 
distance from the Galactic center ($R_{\rm apo} > 25$~kpc).
We apply no explicit orbital eccentricity requirements for 
outer halo membership, although the criteria essentially 
select stars on relatively eccentric orbits since they
are presently in the Solar neighborhood and yet must travel to
great distances. 
In contrast to the inner halo membership, stars only need to possess
one of these characteristics to be included in the outer halo 
population.
While our selection criteria may eliminate 
genuine members of either population, they help to ensure that
there is minimal contamination between the two populations.

Interestingly, our inner halo criteria may allow for the inclusion of
a few thick disk stars on mildly eccentric orbits.
To eliminate these stars, we adopt one additional criterion for 
membership in the inner halo.
Assuming that the stellar thick disk and halo populations 
each have Gaussian velocity distributions in each of the three 
components (e.g., \citealt{bensby03}, eqns.\ 1 and 2), 
we calculate the relative probability that a star is
in the thick disk versus the halo.
We use the velocity dispersions
($\sigma_{U}$, $\sigma_{V}$, $\sigma_{W}$)~=~(46, 50, 35)~\kmsec\ 
for the thick disk and (141, 106, 94)~\kmsec\ for the halo,
as well as the rotation speeds $\langle V_{\phi} \rangle = 200$~\kmsec\ and 
$\langle V_{\phi} \rangle = 40$~\kmsec\ for the thick disk and halo,
respectively \citep{chiba00}.
We require that inner halo members
are at least twice as likely to be members of the halo than the thick disk.
This eliminates less than 15\% of the stars that would otherwise be
classified as members of the inner halo.

Previous investigations \citep{stephens99,stephens02,fulbright02}
have employed very similar criteria to ours for 
defining outer halo membership, including stars with retrograde Galactic
orbits or probing to large Galactocentric distances in and above the 
plane of the disk.  
One unique feature of our study is the adoption of several selection
criteria for defining membership in the inner halo, rather than 
simply identifying stars at less extreme maximum distances as inner halo
members.
This advance has been made possible because of large surveys (e.g.,
SDSS/SEGUE) that have identified very large numbers of metal-poor stars 
in the Solar neighborhood.

One abundance criterion---but not [Fe/H]---is necessary to exclude 
stars whose abundances
do not reflect their initial composition.
When signatures of the slow neutron-capture nucleosynthesis process
(``\spro''; see \S~\ref{ncapstuff}) are enhanced in
metal-poor stars, this is regarded as evidence of past mass transfer
from an undetected companion star that passed through the 
asymptotic giant branch (AGB) phase of stellar evolution.
Since the metal content of these stars has changed over the
course of their lives, we exclude them
membership in either halo population.\footnote{
The pure-$s$-process [Ba/Eu] ratio is predicted to be 
[Ba/Eu]$_{\rm s} = +0.63$ \citep{simmerer04}.
We exclude stars with [Ba/Eu]~$>+0.5$ and [Ba/Fe]~$>+1.0$.
Unfortunately, Eu has been measured in only $\approx$35\% 
of the stars in our entire sample, so it is likely that some stars 
that have experienced \spro\ enrichment during their lifetimes 
will remain in our sample.
Fortunately, though, nucleosynthesis in the envelope of stars during the 
AGB phase has a negligible effect on most light elements, 
(C, N, O, F, Ne, and perhaps Na being notable exceptions; 
e.g., \citealt{straniero06})
including most of those examined in detail in our study.}
Later, in \S~\ref{rspro}, we examine these stars 
for any common kinematic signatures.

The members of both the inner and outer halo populations according
to these definitions are listed, along with the abundance ratios
we consider, in Tables \ref{innerabundtab} and \ref{outerabundtab}.

\subsection{Classification Uncertainties}
\label{classuncertainties}

\begin{figure}
\epsscale{1.15}
\plotone{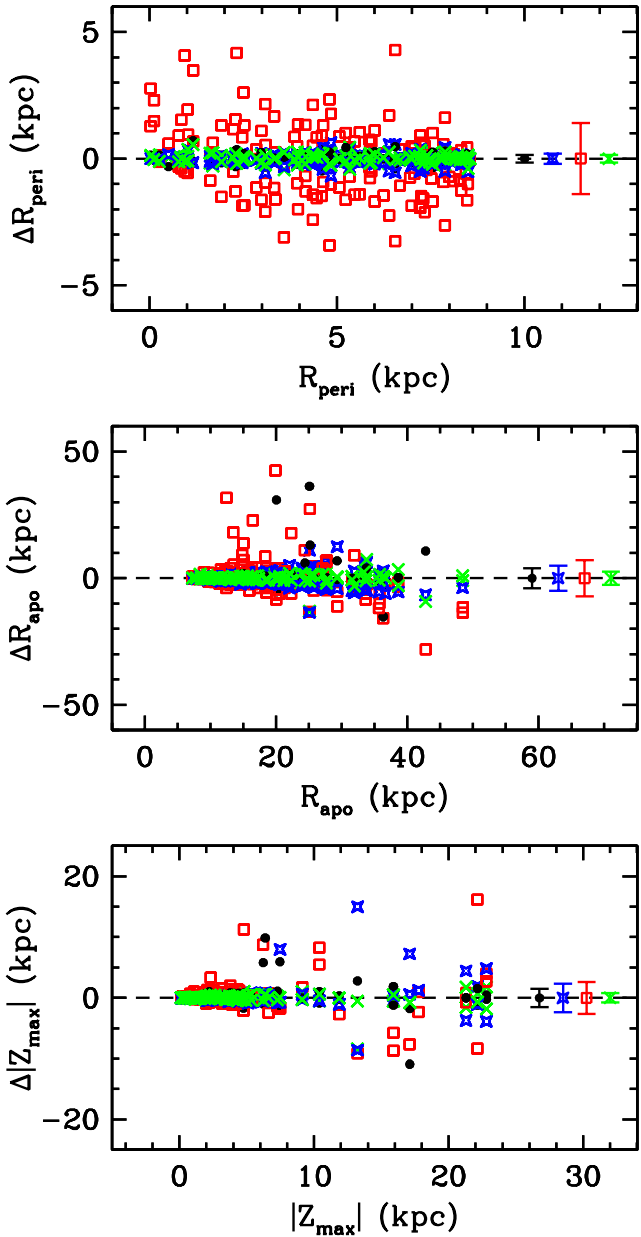}
\caption{
\label{uncertainties}
Uncertainties in the derived orbital parameters.
Changes are shown for alterations in the distance
(dark gray circles, black in the online edition),
the mass of the Galaxy
(open four-point stars, blue in the online edition),
the proper motions (open squares, red in the online edition),
or the radial velocity (light gray ``X'' points, green in the online edition).
The $\Delta$s refer to (modified parameter)$-$(unmodified parameter).
RMS uncertainties are shown for each set of $\Delta$ values
on the right side of each panel.
[Please see the electronic edition for a color version
of this figure.]
}
\end{figure}

In Figure~\ref{uncertainties} we examine the changes in the derived
orbital parameters resulting from variations of the 
input measurements within reasonable uncertainties.
We use the subsample of stars without prior kinematic analysis
(the 309 stars noted in \S~\ref{litdata}) for this test.
The changes are obtained by adjusting the distance by $\pm$20\%,
the proper motions by $\pm$5~mas~yr$^{-1}$ 
(or 20\% for small values of the proper motion), or 
the radial velocity by $\pm10$~\kmsec.
We also examine the effect of changing 
the mass of the Galaxy by $\pm$10\%.
RMS values for each set of changes are shown on the right side of 
each panel in Figure~\ref{uncertainties}.
It is clear that, for most stars, these changes do not introduce 
appreciable uncertainty into the derived orbital parameters.
For a small fraction of stars ($\approx$5\%), the changes are large enough 
that they could cause an
inner halo star to be classified as an outer halo star or vice versa.
While some stars scatter in and
out of the two populations, we confirm that \textit{no} stars 
in this sample actually change their classification from inner to outer
(or vice versa) with these variations in the input parameters.
This result is a consequence of the multiple requirements necessary
for inclusion in the inner halo population and the wide buffers placed
between the kinematic and orbital properties of the two populations.

Confident that our definitions of the inner and outer halo populations
are robust against uncertainties in the measurements, we proceed
to compare the compositions of stars in the inner and outer halo populations.

\section{A Caution: the Connection between Low-Metallicity
Stars and Retrograde Orbits}
\label{caution}

\begin{figure}
\epsscale{1.15}
\plotone{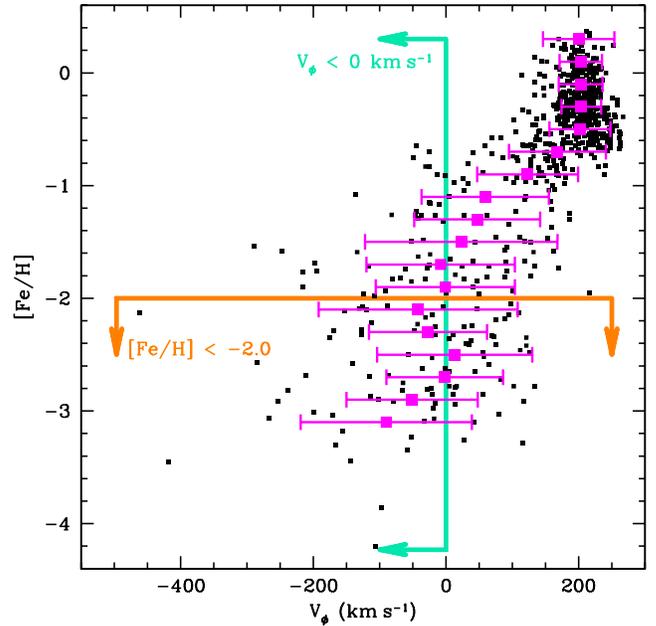}
\caption{
\label{cautionplot}
Rotational velocity $V_{\phi}$ versus [Fe/H] for our sample.
Of the $\sim$700~stars represented on this plot, 94 have retrograde
Galactic orbits ($V_{\phi} < 0$~\kmsec; indicated in turquoise 
in the online edition) and
93 have [Fe/H]~$< -2.0$ (indicated in orange in the online edition).
A large number of stars (53) fit both categories:
57\% of stars with [Fe/H]~$<-2.0$ are on retrograde orbits, and
56\% of stars on retrograde orbits have [Fe/H]~$<-2.0$.
The large gray squares (magenta in the online edition)
indicate $\langle V_{\phi} \rangle$ and the associated
1 standard deviation scatter for stars in 0.2~dex bins of [Fe/H], 
but we emphasize that our sample is biased toward metal-poor stars
with large space velocities, hence the disproportionately large
number of stars with [Fe/H]~$< -2.0$ and significant retrograde
velocities.
[Please see the electronic edition for a color version of this figure.]
}
\end{figure}

Many authors have examined the relationship between 
[Fe/H] and $V_{\phi}$ to characterize the kinematic properties
of the thick disk and halo in an effort to better understand
the formation mechanisms of these components
\citep[e.g.,][]{norris86,sandage87,norris89,beers95,chiba98,chiba00}. 
Our sample is not designed to rederive the relationship between
[Fe/H] and $V_{\phi}$,
but we must assess the effect of the proper motion
bias---which selects against stars with small proper motions---in 
our sample before proceeding.
Examination of the upper left panel of Figure~\ref{innerouterplot}
reveals that many metal-poor stars in our sample have retrograde orbits and
vice-versa.
We show this more explicitly in Figure~\ref{cautionplot}. 
Of the 94 stars in our sample on retrograde orbits, 53
(56\%) also have [Fe/H]~$< -2.0$.
Turning the problem around reveals that, of the 93 stars in our
sample with [Fe/H]~$<-2.0$, 53 (57\%) are on retrograde orbits.
This correlation strengthens at lower metallicities:
35 of the 54 stars (65\%) with [Fe/H]~$<-2.5$ are on retrograde orbits, and
13 of the 17 stars (76\%) with [Fe/H]~$<-3.0$ are on retrograde orbits.

The mean Galactic rotational velocities, 
$\langle V_{\phi} \rangle$, binned in 0.2~dex intervals in [Fe/H],
are shown in Figure~\ref{cautionplot}.
We find an approximate relation between $V_{\phi}$ and [Fe/H] of
$\Delta\langle V_{\phi}\rangle / \Delta$[Fe/H]~$ \approx
-140$~km~s$^{-1}~{\rm dex}^{-1}$ over 
$-2.0 <$~[Fe/H]~$< -0.6$, lessening to 
$\Delta\langle V_{\phi}\rangle / \Delta$[Fe/H]~$ \approx
-90$~km~s$^{-1}~{\rm dex}^{-1}$ over 
$-3.0 <$~[Fe/H]~$< -2.0$.
This relationship is a direct consequence of our
proper motion bias introduced in \S~\ref{litdata},
which would tend to select against very metal-poor stars
with near-zero or prograde net rotation, and 
is not a contradiction to previous results such
as \citet{norris86}, \citet{beers95}, or \citet{chiba00}.
The most recent of these studies found 
$\Delta\langle V_{\phi}\rangle / \Delta$[Fe/H]~$ \approx 
160$~km~s$^{-1}~{\rm dex}^{-1}$ in the range 
$-1.7 <$~[Fe/H]~$<-0.6$ and a nearly constant 
$\Delta\langle V_{\phi}\rangle / \Delta$[Fe/H]~$ \approx
0$~km~s$^{-1}~{\rm dex}^{-1}$ for
$-2.6 <$~[Fe/H]~$<-1.7$ in the Solar neighborhood.
Presumably, if precise proper motions (even small ones) 
were known for all the metal-poor stars found in the literature,
our relationship between $\langle V_{\phi}\rangle$ and [Fe/H]
would flatten out for the same metallicities as found by 
previous studies.

Some metal-poor stars are identified by high proper motion
searches and would be expected to exhibit a kinematic bias;
however, a large percentage of metal-poor stars 
are identified with no kinematic selection criteria
(via objective-prism surveys or ultraviolet excess), 
although these surveys do avoid stars at low Galactic latitude.
Also, investigators performing abundance analyses 
tend to preferentially select the most metal-poor stars for 
high-resolution followup.
Whether because of a selection bias or a true physical preference for
extreme orbits (or both), a significant fraction of the
stars with [Fe/H]~$\lesssim -2.0$ that have been subject to
high-resolution abundance analyses over the last 15 years
have retrograde orbits.
It is possible that the large numbers of very metal-poor stars
with [Fe/H]~$\lesssim -3.0$ found in the last few years could
reintroduce a slope into the relationship between 
$\langle V_{\phi}\rangle$ and [Fe/H] at low metallicities,
and such a reexamination should be undertaken in the near future.

This relation shown in Figure~\ref{cautionplot} 
is important in the context of examining
kinematic and chemical correlations in metal-poor stars.
Any abundance trend that is preferentially found
in low-metallicity stars will also be preferentially found
in stars with retrograde velocities, insofar as those
velocities are computed based on proper motion measurements
with an intrinsic bias.
As such, we caution that any abundance trends correlated with
retrograde orbits alone should not be over-interpreted as 
signatures of past accretion events if they
can be attributed to nucleosynthetic patterns that
inherently occur at low metallicity.

\begin{figure}
\epsscale{1.15}
\plotone{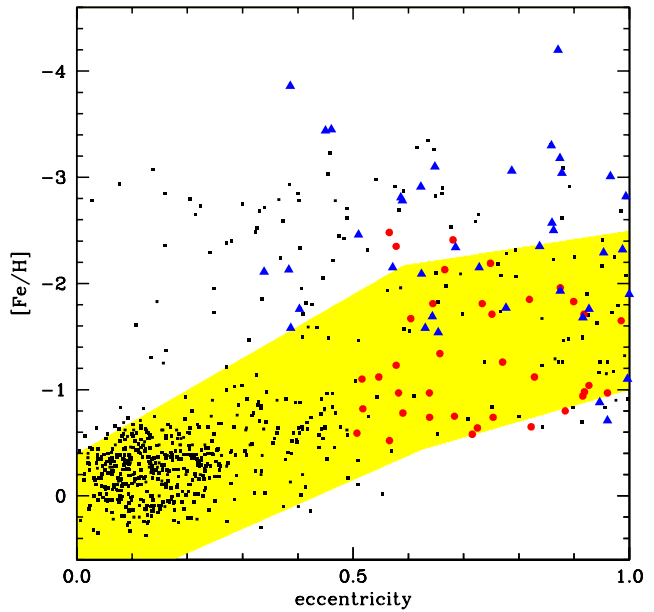}
\caption{
\label{els62plot}
The relationship between metallicity and Galactic orbital eccentricity
in our sample.
Symbols are the same as in Figure~\ref{innerouterplot}.
The shaded band (yellow in the online edition) represents the 
relationship identified in Figure~4 of \citet{eggen62}.
[Please see the electronic edition for a color version of this figure.]
}
\end{figure}

In Figure~\ref{els62plot} we show the relationship between 
metallicity and orbital eccentricity for our inner and outer halo 
populations, similar to Figure~4 of \citet{eggen62}.
The shaded band in this figure represents the locus of stars
used by \citet{eggen62}.  
(We convert ultraviolet excess, $\delta (U-B)$, to [Fe/H] 
using the approximate relationships given in \citealt{sandage87}.)
Their band encompasses the disk stars at high metallicity and low
eccentricity as well as the greater part of our inner halo population,
but their sample did not extend to metallicities much lower than 
[Fe/H]~$= -2.5$.
Many studies since \citet{eggen62} have used larger, unbiased 
datasets to demonstrate 
that there is no correlation between orbital eccentricity and metallicity 
(e.g., \citealt{chiba00} and references therein) in the halo.
While our inner halo is explicitly chosen to include stars with
high eccentricities and our outer halo effectively selects for
these stars, too, Figure~\ref{els62plot} shows that our entire sample
does include metal-poor stars across the full range of eccentricity.

\section{Abundance Results}
\label{results}

\subsection{Trends with $R_{\rm apo}$ and $|Z_{\rm max}|$}
\label{trends}

\begin{figure*}
\epsscale{1.00}
\plotone{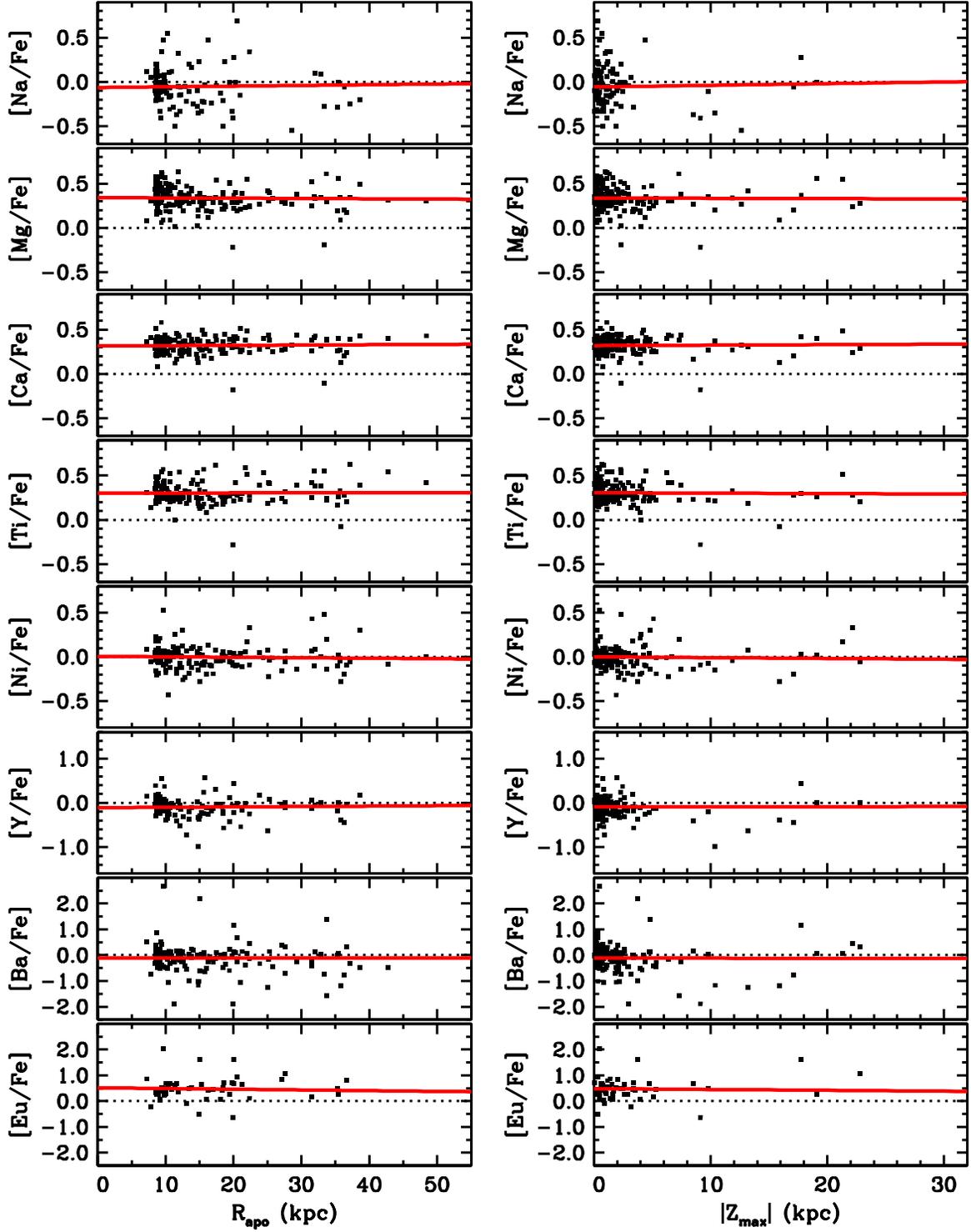}
\caption{
\label{trendplot}
Elemental abundance trends as a function of $R_{\rm apo}$ and 
$|Z_{\rm max}|$ for stars with [Fe/H]~$<-1.0$.
Least squares fits are shown as solid gray lines (red in the online edition).
The solar ratios are indicated by the dotted lines.
[Please see the electronic edition for a color version of this figure.]
}
\end{figure*}

In Figure~\ref{trendplot} we examine the trends of 
the sodium (Na), magnesium (Mg), calcium (Ca), titanium (Ti), 
nickel (Ni), yttrium (Y), barium (Ba), and europium (Eu) to Fe ratios
as a function of $R_{\rm apo}$ and $|Z_{\rm max}|$
for stars with [Fe/H]~$<-1.0$.
Linear least-squares fits are also shown.
No slopes are significant at the 2$\sigma$ level.
Using a sample of 11 stars, 
\citet{stephens99} found no trends between [$\alpha$/Fe] 
and $R_{\rm apo}$ or $|Z_{\rm max}|$, a result which we confirm.
Using a larger sample (56 stars) and a homogeneous abundance analysis,
\citet{stephens02} found a slight decrease in 
$\langle$[$\alpha$/Fe]$\rangle$ 
($-0.0012$~dex~kpc$^{-1}$) from $R_{\rm apo} \sim 8$--100~kpc.
Since they used a homogeneous abundance analysis, we defer to their
result, but we note that their slope only represents a very subtle
change in [$\alpha$/Fe] of $\sim 0.1$~dex over $\sim 90$~kpc.
They also found an increase in the cosmic scatter 
of [Y/Fe] at larger $R_{\rm apo}$.
We propose that the complex nucleosynthetic origins of the 
light neutron-capture species (see \S~\ref{rspro}) must be better 
understood before attempting any serious interpretation of these results.
\citet{stephens02} reported no other dependences of 
abundance ratios with orbital parameters.
Combining the \citet{stephens02} and \citet{fulbright00,fulbright02}
datasets, \citet{fulbright04a} noted slight decreases in all of the light
element ratios with respect to Fe (i.e., Na, Mg, Al, Si) and 
weaker decreases of heavier elements (i.e., Ca, Ti, Ni, Y, and Zr)
when using Galactic rest-frame velocity as a surrogate of the
model-dependent $R_{\rm apo}$.
Our larger---but inhomogeneous---dataset does not show these trends.
The trends reported by \citet{nissen97}---whose dataset differed
significantly with regard to both kinematics and metallicity range
from subsequent studies---have largely not been reproduced by those
studies, including ours.

\subsection{$\alpha$ and Iron-Peak Elements in the Inner and Outer Halo
Populations}
\label{alphafe}

\begin{figure}
\epsscale{1.15}
\plotone{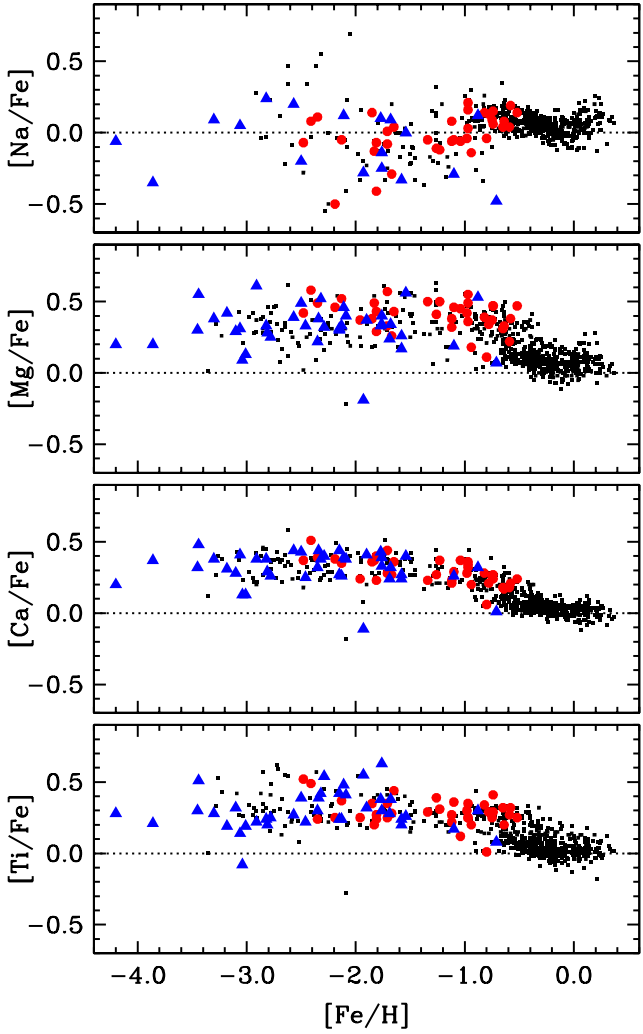}
\caption{
\label{abundplot1}
[Na/Fe], [Mg/Fe], [Ca/Fe], and [Ti/Fe] abundance ratios for our
inner (dark gray circles, red in the online edition) 
and outer (black triangles, blue in the online edition) 
halo populations.
Stars that did not meet the kinematic criteria for these two 
populations are shown as small dots.
[Please see the electronic edition for a color version of this figure.]
}
\end{figure}

\begin{figure}
\epsscale{1.15}
\plotone{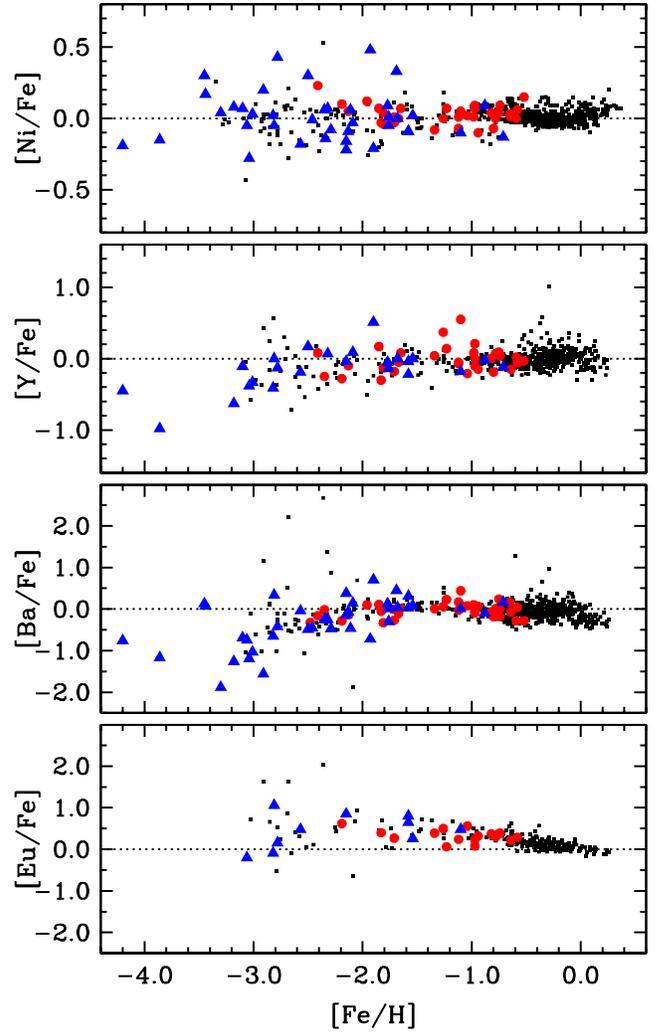}
\caption{
\label{abundplot2}
[Ni/Fe], [Y/Fe], [Ba/Fe], and [Eu/Fe] abundance ratios for our
inner and outer halo populations.
Symbols are the same as in Figure~\ref{abundplot1}.
Upper limits are not displayed.
[Please see the electronic edition for a color version of this figure.]
}
\end{figure}

Figures~\ref{abundplot1} and \ref{abundplot2} display the 
logarithmic abundance ratios of Na, Mg, Ca, Ti, and Ni to Fe
(relative to Solar) for our entire sample 
and for members of the inner and outer halo populations.  
For [Fe/H]~$<-1.0$, 
both the inner and outer halo populations show super-Solar 
[Mg/Fe], [Ca/Fe], and [Ti/Fe] ratios 
(hereafter loosely defined as [$\alpha$/Fe])
scattered around $+0.3$~dex, 
with no apparent slope below [Fe/H]~$\lesssim-1.5$.
\citet{stephens02} reported an increase in [$\alpha$/Fe] with decreasing
[Fe/H], a trend not reproduced elsewhere in studies with large numbers
of very metal-poor stars 
\citep{mcwilliam95b,carretta02,cayrel04,cohen04b,arnone05,lai08}.
The [Mg/Fe] scatter may increase at low metallicities, and
\citet{stephens02} only studied 4 stars with [Fe/H]~$<-3.0$.
[Na/Fe] exhibits significantly larger scatter (increasing scatter with
decreasing metallicity, at least to [Fe/H]~$\sim -2.3$)
than any of the [$\alpha$/Fe] ratios does, 
and [Na/Fe] approximately follows the Solar ratio.
Ni should be produced along with Fe, and [Ni/Fe]
correlates with [Fe/H] at all metallicities with
relatively small scatter (increasing also to a maximum below
[Fe/H]~$\lesssim -2.0$).

One fact is readily apparent from these plots:
the bulk of stars in our inner halo sample only are found
with metallicities in the range 
$-2.5 \lesssim $~[Fe/H]~$\lesssim -0.5$, 
while most stars in our outer halo are found over the 
metallicity range $-3.5 \lesssim $~[Fe/H]~$\lesssim -1.5$.
\citet{carollo07} claimed that the metallicity distribution function
(MDF) of the inner halo peaks around [Fe/H]~$\sim -1.6$ and the 
MDF of the outer halo peaks around [Fe/H]~$\sim -2.2$; our results
support these assertions.
Nevertheless, we caution that (1) our proper motion bias likely 
selects against lower-metallicity stars on prograde orbits and 
(2) our sample is drawn from studies
designed to select interesting metal-poor stars for detailed 
abundance analyses, and therefore it should not be used for
any assessments of metal-poor MDF's.

Our sample is sensitive to chemical differences in the inner and
outer halo populations for stars with the same Fe abundance. 
Between $-2.3 <$~[Fe/H]~$<-1.6$, the mean [Mg/Fe] of the outer halo 
($\langle$[Mg/Fe]$\rangle = 0.30$, 
$\sigma_{\rm mean} = 0.05$)
is slightly lower than the mean [Mg/Fe] of the inner halo
($\langle$[Mg/Fe]$\rangle = 0.40$, 
$\sigma_{\rm mean} = 0.04$), 
even when the well-known $\alpha$-poor star \mbox{G~004-036} 
([Fe/H]~$=-1.93$, $R_{\rm apo} = 33^{+6}_{-4}$~kpc, \citealt{ivans03}) 
is excluded.\footnote{
The quoted uncertainties in the orbital parameters
are computed by the same methods discussed in \S~\ref{classuncertainties}.}
This difference is not obvious in [Ca/Fe] or [Ti/Fe], and the scatter
in [Ca/Fe] for both populations is noticeably smaller than the 
scatter in either [Mg/Fe] or [Ti/Fe].
For all three of [Mg/Fe], [Ca/Fe], and [Ti/Fe], it is worth noting that
the ``extreme'' [X/Fe] ratios at a given [Fe/H] are predominantly
found in the outer halo population.

In Figure~\ref{abundplot2}, this interesting trend begins to emerge 
more clearly with [Ni/Fe]:
the inner halo abundance ratios at a given [Fe/H] appear much more
tightly correlated than the outer halo abundance ratios.
The scatter in inner halo [Ni/Fe] is commensurate with the typical
abundance uncertainties in a given measurement, typically 0.1--0.2~dex,
while the scatter in the outer halo [Ni/Fe] is typically 0.5--0.7~dex
(at least for [Fe/H]~$\lesssim-1.8$).

\begin{figure}
\epsscale{1.15}
\plotone{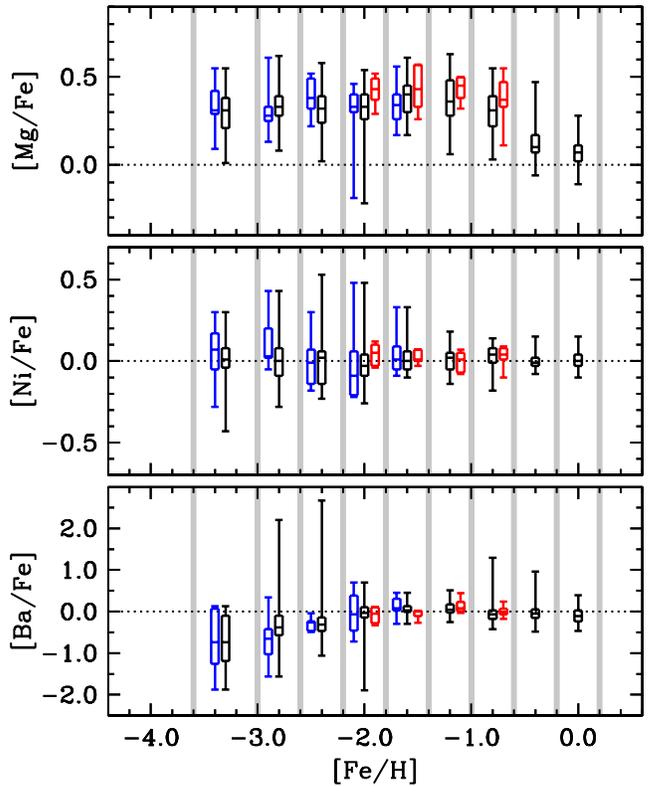}
\caption{
\label{binnedplot}
Binned abundance ratios for [Mg/Fe], [Ni/Fe], and [Ba/Fe],
displayed as quartile boxplots.
Black boxes represent all stars in the bin 
(including the inner and outer halo populations), 
light gray boxes (red in the online edition) 
represent the inner halo stars in the bin,
and dark gray boxes (blue in the online edition)
represent the outer halo stars in the bin.
The divisions in the [Fe/H] bins are indicated by vertical gray stripes.
[Please see the electronic edition for a color version of this figure.]
}
\end{figure}

In Figure~\ref{binnedplot} we display the abundance ratios
for [Mg/Fe] and [Ni/Fe] binned as a function of [Fe/H].
A boxplot is shown for each [Fe/H] bin (typically 0.4~dex wide), 
displaying the median, inner quartiles, and extremes of 
the entire sample, only the inner halo stars, and only the outer halo stars.
For $-2.2 \lesssim$~[Fe/H]~$\lesssim -1.4$ (where 
there is significant abundance overlap between the two populations),
the median [Mg/Fe] ratios for the inner halo are consistently higher than the
outer halo medians by 0.10--0.20~dex. 
The mean [Ni/Fe] ratios of both populations trace the Solar ratio very closely.
For [Ni/Fe], it is apparent that the extremes of the
inner halo are significantly smaller than the extremes of the outer halo.
In this case we point to the different degrees of scatter in the 
inner versus outer populations 
as evidence for chemical differences between the inner and outer 
halo populations;
no stars with significant deviations from the median ratios
are members of the inner halo population.
The precise amount of scatter of each population may be affected
by the inhomogeneous nature of our sample, but the relative 
scatter between the two populations is robust. 

\begin{figure*}
\epsscale{1.00}
\plotone{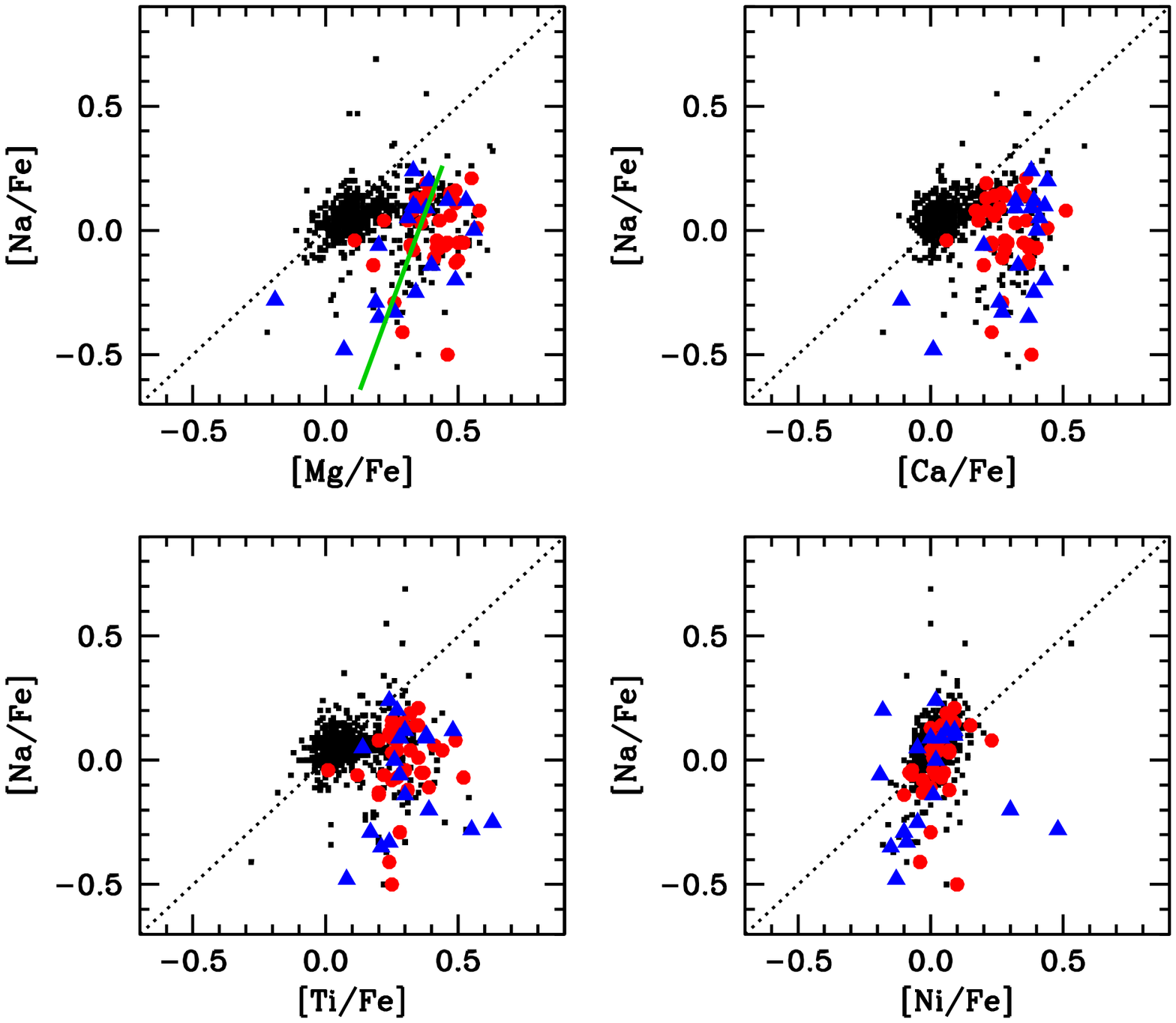}
\caption{
\label{naplots}
[Na/Fe] as a function of [Mg/Fe], [Ca/Fe], [Ti/Fe], and [Ni/Fe]
for our inner and outer halo populations.
Symbols are the same as in Figure~\ref{abundplot1}.
The dotted line represents [X/Na]~$= +0.0$.
The solid line (green in the online edition)
in the top left panel is the correlation between
[Mg/Fe] and [Na/Fe] reported by \citet{hanson98}.
[Please see the electronic edition of the journal for a color version
of this figure.]
}
\end{figure*}

Our [Na/Fe] ratios in Figure~\ref{abundplot1}
exhibit a large degree of scatter.
Two of our literature sources for low metallicity stars,
\citet{honda04a,honda04b} and \citet{barklem05}, did not report 
[Na/Fe] ratios for their samples, thus at low
metallicities ([Fe/H]~$\lesssim -2.0$) there are not enough 
measurements to adequately compare the inner and outer halo populations.
In Figure~\ref{naplots}, we show [Na/Fe] as a function of
[Mg/Fe], [Ca/Fe], and [Ti/Fe].
The line in the upper left panel of this figure shows the
correlation between [Mg/Fe] and [Na/Fe] found in field halo giants
by \citet{hanson98}, which generally matches our metal-poor sample.
A similar trend (not shown) exists between [Ca/Fe] or [Ti/Fe] and 
[Na/Fe]. 
Outer halo stars generally occupy the extremes of each distribution,
particularly the Na-depleted extremes.
This implies that these stars formed from---at least in 
part---an incompletely mixed ISM where the yields of 
individual Type~II SNe events could still be ``noticed'' 
against the overall chemical background of the ISM.

\citet{hanson98} reported both 
an increase in the [Na/Fe] scatter with decreasing [Fe/H]
and an increase in the [Na/Fe] scatter when comparing stars on
retrograde orbits to those on prograde orbits.
To investigate this trend, we turn to a species with more 
measurements in our dataset, Ni.
Previous analyses have revealed correlations between
[Na/Fe] and [Ni/Fe] ratios and stellar kinematic properties
(e.g., \citealt{nissen97}, \citealt{hanson98}, 
\citealt{shetrone03}, \citealt{venn04}).
We show this relationship in our sample in Figure~\ref{naplots}.
The Na-Ni relationship originates from the neutron-rich nature
of the dominant isotopes of these species, which can be produced
in Type~II SNe, albeit in non-Solar ratios.
(See \citealt{venn04} for an extensive discussion of
the Na-Ni nucleosynthesis relationship in this context.)
This relationship breaks down in material
enriched by Type~Ia SNe products, since 
Type~Ia SNe produce very little $^{23}$Na, 
so we only examine stars with [Fe/H]~$<-1.0$.
Ni-poor stars ([Ni/Fe]~$\leq-0.2$)
are preferentially associated with stars on retrograde orbits; however,
Ni-rich stars ([Ni/Fe]~$\geq +0.2$) and Ni-normal stars
($-0.05 \leq$~[Ni/Fe]~$\leq +0.05$) also exhibit a preference 
for retrograde or no net rotation orbits in our sample.
The same effect occurs even if we only consider stars with 
[Fe/H]~$<-2.0$, where our proper motion bias is most pronounced.
Thus we are unable to confirm or refute the correlation between
non-Solar [Ni/Fe] (and, by extension, {Na/Fe]) ratios and retrograde orbits.

\subsection{Neutron-Capture Species in the Inner and Outer Halo Populations}
\label{ncapstuff}

Nuclei heavier than the iron group are formed by the addition of 
neutrons to existing seed nuclei.
The timescales for neutron- ($n$-) captures determine the resulting
abundance patterns.
If the average time between successive 
neutron captures is less than the typical halflife against $\beta^{-}$ 
decay, this is referred to as the slow- ($s$-) \ncap\ process.
In contrast, the rapid- ($r$-) \ncap\ process occurs when many neutrons 
are added before any $\beta^{-}$ decays can occur.
These two processes result in very different heavy element 
abundance patterns.
An exact site for the \rpro\ has yet to be conclusively identified,
but the short stellar timescale necessary to produce \rpro\ enrichment
in stars with [Fe/H]~$\lesssim -3.0$ suggests that an association 
with massive core-collapse Type~II SNe is likely.
Low and intermediate mass ($\sim$~1.5--3.0~$M_{\odot}$) stars that pass
through the AGB phase of evolution are 
the primary source of \spro\ material \citep[e.g.,][]{busso99,straniero06}. 

In Figure~\ref{abundplot2} we examine the abundances of the \ncap\ 
species [Y/Fe], [Ba/Fe], and [Eu/Fe] in our inner and outer
halo populations.
These ratios display very little scatter for [Fe/H]~$\gtrsim-1.8$,
but they exhibit considerable scatter below [Fe/H]~$\lesssim-1.8$.
A familiar trend reappears:
the inner halo [X/Fe] ratios follow (roughly) a monotonic relationship
with [Fe/H], while the outer halo [X/Fe] ratios scatter about 
appreciably at a given [Fe/H].
For example, the scatter in inner halo [Ba/Fe] is larger ($\sim$~0.5~dex)
than the typical measurement uncertainty ($\sim$~0.1--0.2~dex)
and follows the Solar ratio, 
yet the outer halo [Ba/Fe] scatter is \textit{much} larger, 
typically 1--2~dex 
(see also, e.g., \citealt{mcwilliam98} and \citealt{francois07}).
This pattern may also be observed with [Y/Fe], though the outer halo
scatter is much less extreme.
It is more difficult to discern these trends with [Eu/Fe], 
for which fewer measurements exist, and 
these effects may also be attributed only to increased [Eu/Fe] scatter 
at the lowest metallicities.
The [Ba/Fe] ratios, binned by [Fe/H], are also shown in 
Figure~\ref{binnedplot}.
The mean [Ba/Fe] of all stars traces the Solar value over the range
$-2.0 \lesssim$~[Fe/H]~$\lesssim +0.0$, only declining at the lowest 
metallicities.
The majority of [Ba/Fe] ratios in both populations are similar.
Stars with extreme ratios only comprise the outer halo;
the inner halo ratios are remarkably similar to one another.

The top panel of Figure~\ref{innerouterncap} displays the [Ba/Y] ratio
as a function of [Fe/H] for our inner and outer halo samples.
$^{89}$Y ($Z=39$) contains 50 neutrons, which is one of the 
magic neutron numbers that correspond to closed nuclear shells
and significantly lower the nuclear cross section to further neutron capture.
During the slow ($s$) nucleosynthesis reaction, this bottleneck causes 
lots of nuclei with 50 neutrons to be produced.
Thus Y is representative of the atomic species produced at this first
abundance peak in the \spro.
$^{138}$Ba ($Z=56$) is the dominant isotope of Ba produced in \spro\ 
nucleosynthesis and also contains a magic number of neutrons, 82.
Ba is representative of the atomic species produced at the second abundance
peak in the \spro.
The [Ba/Y] ratio is useful as a probe of the relative amounts of material
produced at these two peaks in the \spro.
The general decline in [Ba/Y] at [Fe/H]~$\lesssim -2.0$ results 
from decreasing Ba contributions from the main \spro\ (see the 
bottom panel of this figure), while Y production from an apparently
primary (i.e., not metallicity-dependent) process remains 
approximately constant to the lowest metallicities observed
(see, e.g., \citealt{travaglio04}).
The inner and outer halo members appear to be randomly 
distributed among the normal scatter for a given [Fe/H], and there
is no obvious correlation with these populations.
(The greatest scatter does appear 
in the outer halo population at [Fe/H]~$\lesssim -2.0$ 
where there is an overall lack of inner halo stars.)

\begin{figure}
\epsscale{1.15}
\plotone{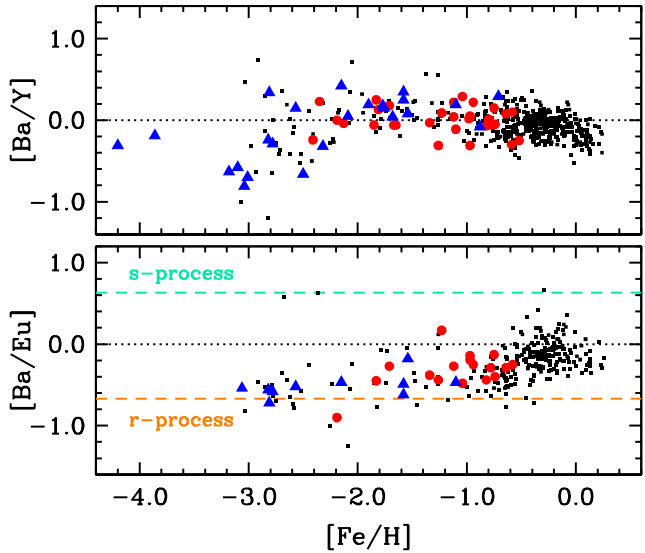}
\caption{
\label{innerouterncap}
[Ba/Y] and [Ba/Eu] abundance ratios for our
inner and outer halo populations.
Symbols are the same as in Figure~\ref{abundplot1}.
The lower gray dashed line (orange in the online edition)
represents the pure \rpro\ ratio and
the upper gray dashed line (turquoise in the online edition)
represents the pure \spro\ ratio as predicted by \citet{simmerer04}.
[Please see the electronic edition for a color version of this figure.]
}
\end{figure}

The bottom panel of Figure~\ref{innerouterncap} displays the [Ba/Eu]
ratio as a function of [Fe/H] for our inner and outer halo samples.
Relative to Ba, very little Eu is produced via the \spro, yet 
Eu is relatively easy to produce in the rapid ($r$) nucleosynthesis
reaction, so the [Ba/Eu] ratio provides a good assessment of the 
relative amounts of $s$- and \rpro\ material present in a star.
We also show the [Ba/Eu] ratios predicted for pure $s$- or \rpro\ 
nucleosynthesis \citep{simmerer04}.
(Recall that we have excluded stars with a pure-$s$-process
signature from membership in our inner and outer halo populations.)
The stellar data generally decline from the Solar [Ba/Eu] ratio at high 
metallicity toward an \rpro\ dominant ratio at low metallicity, though
a small number of stars at low-metallicity show evidence of 
$s$-only enrichment.
Again, the inner and outer halo members appear to be randomly 
distributed among the scatter of all stars at a given [Fe/H].

\section{Discussion}
\label{discussion}

\subsection{The Kinematically and Chemically Diverse Outer Halo}

Is it possible that the inner halo population consistently exhibits
a smaller degree of scatter than the outer halo population because of
our classification process and not an astrophysical phenomenon?
In other words, have we more precisely determined a kinematic population
with our ``and'' selection criterion for the inner halo than with our
``or'' selection criterion for the outer halo?
To address this possibility, we more closely analyze each of our 
selection criteria for the outer halo population.
In Figure~\ref{retroplot} we display the [Mg/Fe] and [Ba/Fe] ratios
of stars on increasingly retrograde Galactic orbital velocities,
in Figure~\ref{highzplot} we display these ratios for stars
with increasingly higher values of $|Z_{\rm max}|$, and 
in Figure~\ref{highrplot} we display these ratios for stars
with increasingly larger values of $R_{\rm apo}$.
Appreciable increases in the [Ba/Fe] scatter are not obvious 
for stars with the most retrograde velocities,
highest distances from the Galactic plane,
or increasing maximum distance from the Galactic center.
Discernible changes in the [Mg/Fe] ratios are not apparent either.
Any set of stars sharing one of these outer halo defining characteristics 
likely would exhibit considerably more abundance 
scatter than the stars in our inner halo population.
The stars in our outer halo population
appear to be genuinely kinematically and chemically uncorrelated.

\begin{figure}
\epsscale{1.15}
\plotone{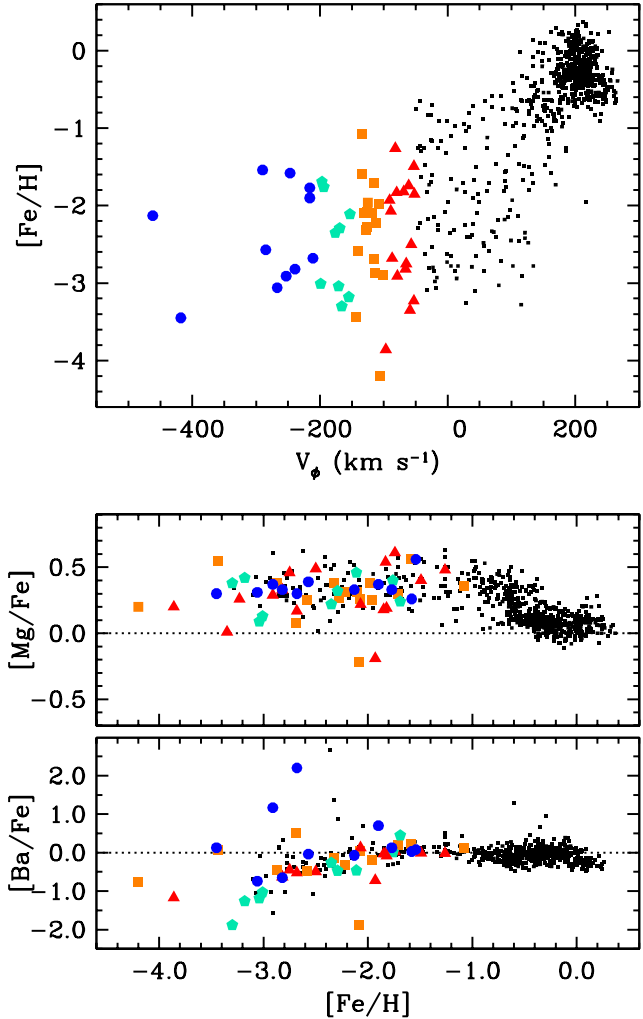}
\caption{
\label{retroplot}
Identifying stars of increasing retrograde rotation about the Galaxy
and associated [Mg/Fe] and [Ba/Fe] abundance ratios.
Stars are selected based only on their rotation velocity:
$-100 < V_{\phi} \leq -50$~\kmsec\ 
(filled gray triangles, red in the online edition),
$-150 < V_{\phi} \leq -100$ 
(filled gray squares, orange in the online edition),
$-200 < V_{\phi} \leq -150$ 
(filled gray pentagons, turquoise in the online edition),
and    $V_{\phi} \leq -200$ 
(filled gray circles, blue in the online edition).
The stars selected according to these definitions are highlighted in 
the lower two panels.
[Please see the electronic edition for a color version of this figure.]
}
\end{figure}

\begin{figure}
\epsscale{1.15}
\plotone{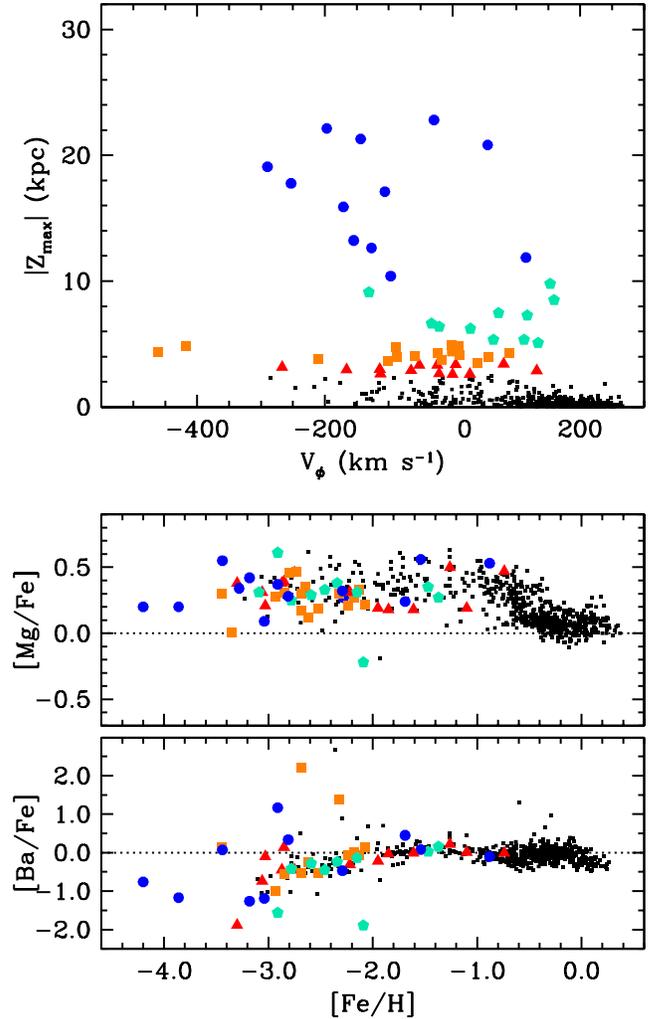}
\caption{
\label{highzplot}
Identifying stars of increasing vertical distance from the Galactic plane
and associated [Mg/Fe] and [Ba/Fe] abundance ratios.
Stars are selected based only on $|Z_{\rm max}|$:
$ 2.5 < |Z_{\rm max}| \leq  3.5$~kpc 
(filled gray triangles, red in the online edition),
$ 3.5 < |Z_{\rm max}| \leq  5$ 
(filled gray squares, orange in the online edition),
$ 5 < |Z_{\rm max}| \leq 10$ 
(filled gray pentagons, turquoise in the online edition),
and  $|Z_{\rm max}| >    10$ 
(filled gray circles, blue in the online edition).
The stars selected according to these definitions are highlighted in
the lower two panels.
[Please see the electronic edition for a color version of this figure.]
}
\end{figure}

\begin{figure}
\epsscale{1.15}
\plotone{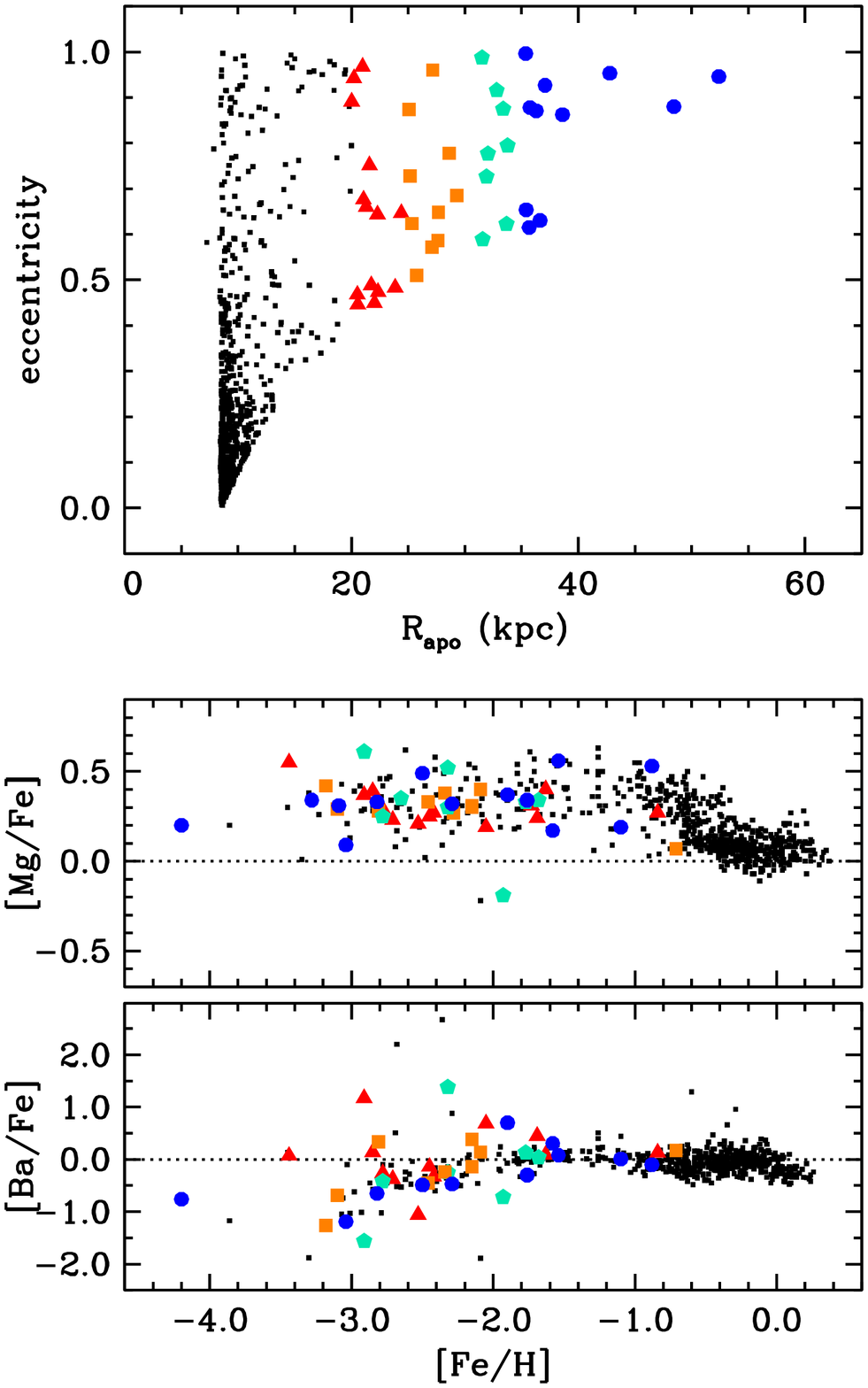}
\caption{
\label{highrplot}
Identifying stars of increasing maximum radial distance from the 
Galactic center and associated [Mg/Fe] and [Ba/Fe] abundance ratios.
Stars are selected based only on $R_{\rm apo}$:
$20 < R_{\rm apo} \leq 25$~kpc 
(filled gray triangles, red in the online edition),
$25 < R_{\rm apo} \leq 30$ 
(filled gray squares, orange in the online edition),
$30 < R_{\rm apo} \leq 35$ 
(filled gray pentagons, turquoise in the online edition),
and  $R_{\rm apo} > 35$ 
(filled gray circles, blue in the online edition).
The stars selected according to these definitions are highlighted in
the lower two panels.
[Please see the electronic edition for a color version of this figure.]
}
\end{figure}

It is somewhat surprising that even a couple of stars in our outer halo
population are found at such a high metallicity with ``standard'' 
elemental abundance ratios for their metallicity.
These stars, HIP~19814 and HIP~117041, with metallicities 
[Fe/H]~$= -0.71$ \citep{stephens02} and $-$0.88 \citep{fulbright00}, 
are on orbits extending 
to $R_{\rm apo} = 27^{+4}_{-3}$~kpc and $52^{+9}_{-1}$~kpc, respectively.
These stars are also on highly eccentric orbits 
($e = 0.96$ and $e = 0.95$, respectively) and
may be in the metal-rich end of the (outer) halo MDF.
The uncertainty in $R_{\rm apo}$ for HIP~19814 could marginally demote this
star from outer halo membership; even so, its Galactic orbit would
remain eccentric.

It has long been common practice to 
assume that ancient metal-poor stars do not accrete any appreciable
amount of metals (certainly not enough to enrich a metal-free star
to [Fe/H]~$\sim -4.0$)
from passage through the gas-rich Galactic 
disk (see commentary on this subject by, e.g., 
\citealt{yoshii81}, \citealt{iben83}, and \citealt{frebel08b}.). 
If this is so, 
then the composition and Galactic orbits of these stars suggest that 
significant metal enrichment ($\sim 1/5$ to $1/8$ Solar Fe) may have
occurred in some localized regions far from the present Galactic disk.
This is in qualitative agreement (but perhaps not quantitative, 
since our local sample of stars may not be representative of the 
bulk of the stellar halo; see \S~\ref{kfa})
with the halo chemical evolution model
presented by \citet{tumlinson06}, who found that some stars in 
the metal-rich end of the halo MDF were forming within the first
few hundred million years after star formation began.

A wide diversity of stellar orbits and chemical compositions is
found in our outer halo population, which is strong evidence that
a significant fraction of the halo was formed from the conglomeration
of small fragments representing a variety of nucleosynthetic
enrichment scenarios \citep[e.g.,][]{searle78}.

\subsection{Relationship to the Inner and Outer Halo Globular Clusters}
\label{globular}

Globular clusters can be classified according to their Galactic
orbital parameters, traditionally defined such that 
``outer halo'' clusters have orbits that take them to much greater
radii from the Galactic center than ``inner halo'' clusters.
Might we learn any additional information by classifying clusters
according to the inner and outer halo population definitions given
in \S~\ref{halokinematics}?
We use a sample of 25 globular clusters with measured distances
and space velocities compiled from the literature by \citet{pritzl05},
who derived mean abundance ratios for each cluster from recent
high-resolution spectroscopic analyses of individual stars in each cluster.
The cluster positions are taken from the most recent version
of the \citet{harris96} catalog (February 2003), and the cluster 
velocities are taken from a series of papers by 
\citet{dinescu99,dinescu00,dinescu01,dinescu03}.
We compute orbital parameters using our model for the Galactic potential.
Table~\ref{gctab} displays the adopted distances and velocities
and our derived orbital parameters for this set of clusters.
Four clusters match our inner halo kinematic criteria 
(M4, M71, NGC~6397, and NGC~6752), ranging in metallicity from
$-2.0 \leq$~[Fe/H]~$\leq -0.7$.
Eight clusters match our outer halo kinematic criteria
(M3, M30, M68, NGC~288, NGC~362, NGC~5466, Pal~5, and Pal~12),
ranging in metallicity from $-2.4 \leq$~[Fe/H]~$-0.7$.
Pal~12 has been conclusively identified as having been stripped
from the Sagittarius dSph;
excluding Pal~12, our outer halo 
globular clusters span the metallicity range $-2.4 \leq$~[Fe/H]~$-1.3$.

In Figure~\ref{globabund1} we compare the [Mg/Fe], [Ca/Fe], and
[Ti/Fe] abundance ratios for our inner and outer halo globular clusters
and our inner and outer halo field stars. 
In all three cases, the abundance ratios of the 
inner halo globular clusters obey the same trends and degree of scatter
defined by the inner halo field stars.
The outer halo globular clusters likewise follow the trends and scatter
of the outer halo field stars, with a few exceptions.
[Mg/Fe] and [Ca/Fe] are marginally low (but still super-Solar)
in NGC~5466, though they
have been derived from a single Cepheid variable \citep{mccarthy97} 
and should be treated with some caution; even so, NGC~5466 is 
within the scatter expected for an outer halo cluster.
[Ti/Fe] is Solar in M68, which helped lead \citet{lee05} to postulate
that M68 may have sampled an IMF biased toward higher masses, where the
Mg and Si overabundances were produced by Type~II SNe.
In this scenario Ti would be primarily produced by by 
lower mass SN~Ia along with the Fe-peak elements.
(See also the extensive discussion in \citealt{lee02}, 
who identified constant [Ca/Fe] but 
decreasing [Si/Fe] and increasing [Ti/Fe] ratios with 
\textit{current} Galactocentric radius in their sample, and
\citealt{gratton04}, who also reproduced this result when using 
$R_{\rm apo}$.)
Finally, M68 does not appear to be associated with the Canis Major
dSph \citep{pritzl05}, as had been suggested by earlier
models \citep{martin04}; however, \citet{pritzl05} have suggested
that it may have an extragalactic origin based on its younger
age, high [Si/Ti] ratio, and high prograde rotational velocity.

\begin{figure}
\epsscale{1.15}
\plotone{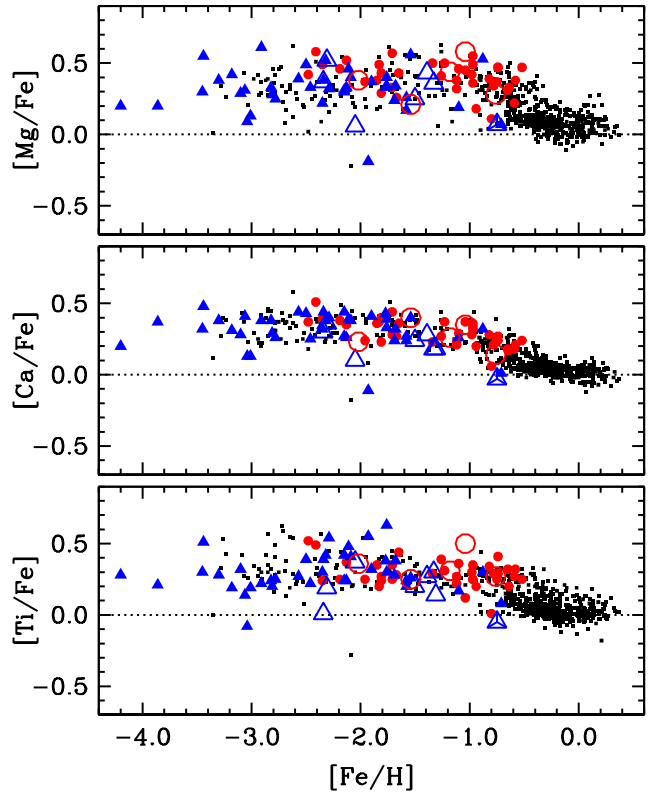}
\caption{
\label{globabund1}
Mean [Mg/Fe], [Ca/Fe], and [Ti/Fe] abundance ratios for globular 
clusters with kinematics like those that define our
inner and outer halo (field star) populations.
Globular clusters with kinematics like the inner halo 
are indicated by the large, open gray circles (red in the online edition), and 
globular clusters with kinematics like the outer halo
are indicated by the large, open black triangles (blue in the online edition).
The triangle marked with an ``X'' indicates Pal~12,
which has been conclusively identified as a cluster
accreted from the Sagittarius dSph.
All other symbols are the same as in Figure~\ref{abundplot1}.
The Solar ratios are indicated by the dotted lines.
[Please see the electronic edition for a color version of this figure.]
}
\end{figure}

Figure~\ref{globabund2} displays the [Y/Fe], [Ba/Fe], and [Eu/Fe]
ratios for our inner and outer halo field stars and globular clusters.
The inner \textit{and} outer halo globular clusters possess 
[Ba/Fe] and [Eu/Fe] ratios that very closely follow these ratios in
the inner halo field stars with very small scatter.
To some degree, this reflects the fact that we have represented
the abundance ratios by means rather than the scatter intrinsic
from one star to another within a given cluster; however,
this scatter is much smaller (typically $\lesssim 0.5$~dex) 
than that found for field stars ($\gtrsim 2$~dex), so it cannot
tell the full story (cf., e.g., \citealt{sneden97,sneden00},
\citealt{ivans01}, \citealt{gratton04}, \citealt{yong08}).
This theme---also present in recent reviews of globular cluster
abundances \citep{gratton04,sneden04}---suggests 
that these globular cluster stars formed from a 
homogenized ISM much like the field stars of the inner halo.
In this sense, the abundance trends traced by our inner halo population
and the globular cluster population may represent a time-averaged
set of chemical yields for a metal-poor stellar population.
If the earliest generations of stars 
pre-enriched the ISM from which the present stars formed, 
this would also explain the lack of stars with
[Fe/H]~$\lesssim -2.5$ in the inner halo and globular clusters.

\begin{figure}
\epsscale{1.15}
\plotone{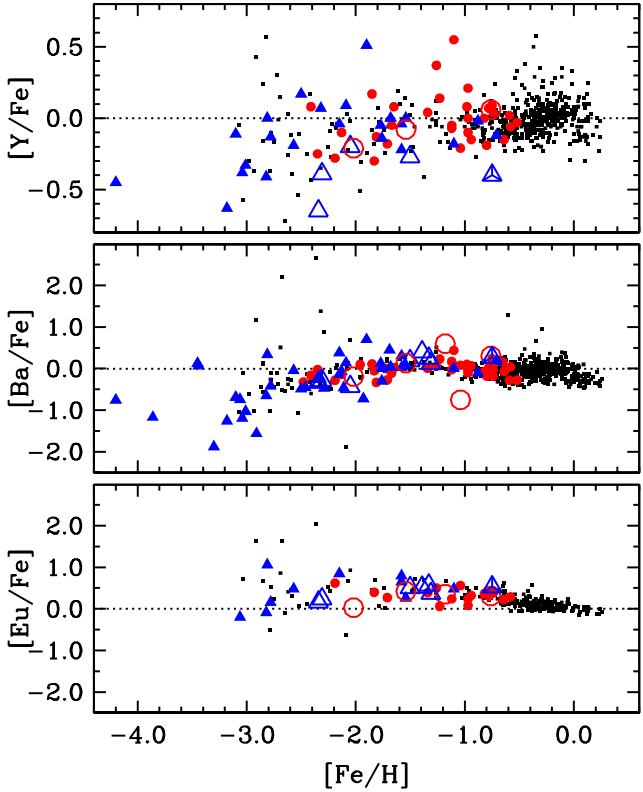}
\caption{
\label{globabund2}
Mean [Y/Fe], [Ba/Fe], and [Eu/Fe] abundance ratios for globular 
clusters with kinematics like those that define our
inner and outer halo populations.
Symbols are the same as in Figure~\ref{globabund1}.
[Please see the electronic edition for a color version of this figure.]
}
\end{figure}

The [Y/Fe] ratios implore us to exercise some caution with this
interpretation.
Mean [Y/Fe] ratios for the five outer halo clusters with kinematic and
abundance data are consistently lower than the mean [Y/Fe] ratios for 
the three inner halo clusters
($\langle$[Y/Fe]$\rangle _{\rm outer} = -0.38$, $\sigma_{\rm mean} = 0.13$; 
$\langle$[Y/Fe]$\rangle _{\rm inner} = -0.08$, $\sigma_{\rm mean} = 0.11$)
These differences are less apparent in Figure~\ref{globabund3}, 
where we display the [Ba/Y] and [Ba/Eu] ratios, yet four of the five
outer halo clusters do have super-Solar [Ba/Y] ratios.
This might suggest a decreased contribution from very massive
SNe (see \S~\ref{rspro}), 
although analysis of a more homogeneous dataset including
abundances for more clusters with kinematic information would be necessary
to draw any robust conclusions.

\begin{figure}
\epsscale{1.15}
\plotone{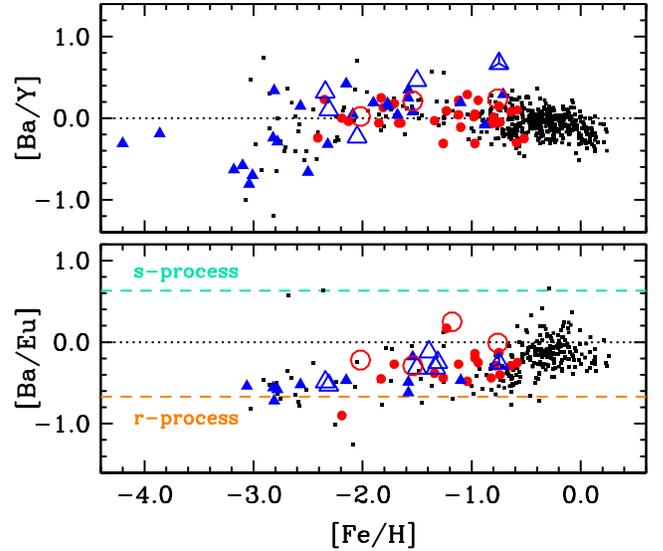}
\caption{
\label{globabund3}
Mean [Ba/Y] and [Ba/Eu] abundance ratios for globular 
clusters with kinematics like those that define our
inner and outer halo populations.
Symbols are the same as in Figure~\ref{globabund1}.
[Please see the electronic edition for a color version of this figure.]
}
\end{figure}

Finally, in Figure~\ref{globtrends} we plot the [Ca/Fe], [Ti/Fe],
[Y/Fe], and [Ba/Fe] ratios as a function of $R_{\rm apo}$, 
$|Z_{\rm max}|$, and $V_{\phi}$.
Excluding Pal~12, [Ca/Fe] exhibits no 
trends with these kinematic properties, while earlier reports of 
a [Ti/Fe] trend with $R_{\rm apo}$ appear less secure with this dataset.
No obvious trends of [Ba/Fe] with kinematics are visible.
Only Y-deficient ([Y/Fe]~$<-0.2$) clusters are found at 
$R_{\rm apo} > 10$~kpc or $|Z_{\rm max}| > 5$~kpc, including Pal~12.
Pal~12 displays the low [$\alpha$/Fe] ratios and extreme kinematics
associated with being a captured cluster.\footnote{
Pal~12 has low Mg, Ca, and Ti, which closely correspond to the 
notoriously low Mg, Ca, and Ti abundances of the Sagittarius dSph
and its associated globular cluster system 
\citep{bonifacio03,tautvaisiene04,cohen04a,sbordone07},
although the low-metallicity globular clusters associated with 
the Sagittarius dSph have elevated [$\alpha$/Fe] ratios 
\citep{mottini08}.}
Could these properties be used to diagnose additional clusters that have 
been captured from dSphs?
Given the inhomogeneous abundance analysis methods, 
wide range of number of stars examined in each cluster,
abundance dispersions within clusters,
and range of mean elemental abundance ratios for clusters already examined,
we suspect that these diagnostics may be more useful for confirming
capture proposals based on kinematic evidence rather than as 
a priori criteria (such as the case made for Pal~12 by \citealt{cohen04a}).

\begin{figure*}
\epsscale{1.00}
\plotone{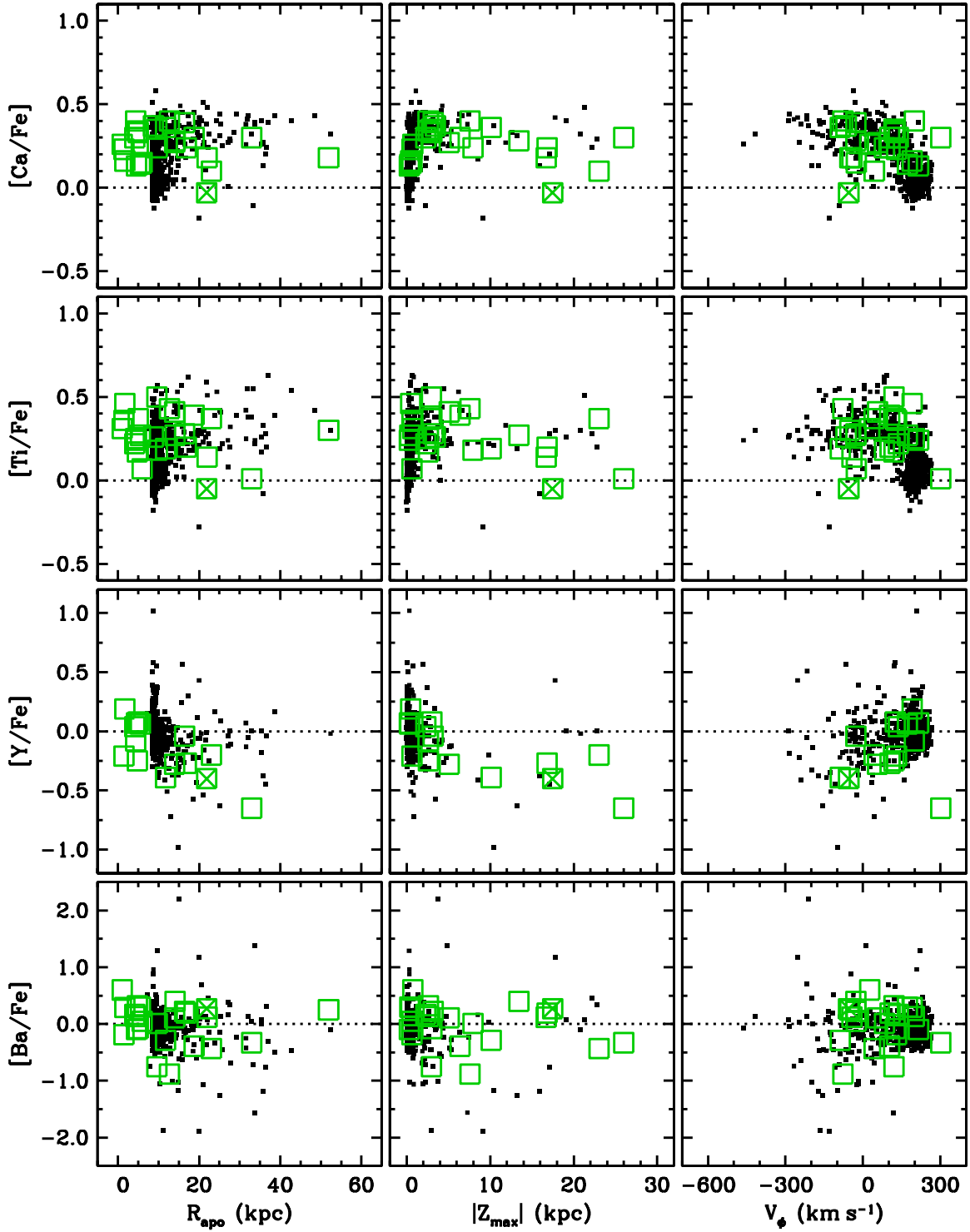}
\caption{
\label{globtrends}
[Ca/Fe], [Ti/Fe], [Y/Fe], and [Ba/Fe] ratios for globular clusters as a 
function of $R_{\rm apo}$, $|Z_{\rm max}|$, and $V_{\phi}$.
Globular clusters are indicated by the large, open black
squares (green in the online edition)
and our stellar sample is indicated by the small gray dots.
Pal~12 is indicated by the square marked with an ``X''.
The Solar ratios are indicated by the dotted lines.
[Please see the electronic edition for a color version of this figure.]
}
\end{figure*}

\subsection{Is the Inner Halo a Remnant of an Accreted Dwarf Galaxy?}

The orbital characteristics we have used to identify members of 
the inner halo population are reminiscent of a clump of stars
found with $e \sim 0.9$, [Fe/H]~$\sim -1.7$,
and near-zero (or perhaps just slightly prograde) net rotation
by \citet{chiba00}.
They speculated that a significant fraction of the stars in 
this clump may have formed from infalling gas with this metallicity.
The very small abundance scatter for other species at a given [Fe/H] 
in our inner halo population supports this interpretation.
\citet{dinescu02} articulated that a slight retrograde
clump of stars spanning $-2.0<$~[Fe/H]~$-1.5$ in the \citet{beers00}
data (employed also by \citealt{chiba00}) could be reconciled
with stellar debris associated with $\omega$~Centauri,
the stripped core of an accreted dSph \citep[e.g.,][]{norris97,dinescu99}
that has a similar peak in its MDF \citep{norris96}.
Could our inner halo population share a similar origin with this 
$\omega$~Cen debris
(which could also help to place age constraints on the inner halo)?
Comparing the chemical abundances of $\omega$~Cen giants \citep{norris95}
with our inner halo sample reveals notable differences in 
Ca, Ti, Ni, Y, and Ba trends with metallicity, 
effectively dismissing this hypothesis.
Furthermore, it is unlikely that a sizable fraction
of the entire stellar halo of the Milky Way Galaxy (which 
we associate with the inner halo population) was composed of
stars accreted from a single dSph.

The inner halo abundances might still suggest that
this population resembles a large-scale stellar stream
\citep[e.g.,][and references therein]{kepley07,helmi08}.
While the spatial density distribution of stellar streams has long
since dissipated, their velocity space density remains more intact
\citep{helmi99}.
In Figure~\ref{angmomplot} we show our inner and outer halo populations
expressed in terms of their angular momenta 
$J_{X} = yW - Vz$, $J_{Y} = zU - Wx$, $J_{Z} = xV - Uy$, and
$J_{\perp} = (J_{X}^{2} + J_{Y}^{2})^{1/2}$ 
(per unit mass, 
where [$x$, $y$, $z$] and [$U$, $V$, $W$] are the positions
and velocities, respectively, of a star in a left-handed 
Galactocentric frame).
The components of the angular momentum for each star are 
also listed in Table~\ref{startab}.
The outer halo stars are preferentially found (as per our 
selection criteria) at low values of $J_{Z}$ and high values of
$J_{\perp}$.
In contrast to the outer halo,
the inner halo stars occupy a limited region of the diagram, and
this region appears as part of the continuous progression from 
stars on disk-like orbits to stars on orbits that take them far
from the Galactic center.  
In this regard the inner halo certainly looks like a more coherent
kinematic component of the halo 
(even if it is too large to be realistically
classified as a stellar ``stream'') than the outer halo, which 
looks like a smattering of stars on completely unrelated orbits.

\begin{figure}
\epsscale{1.15}
\plotone{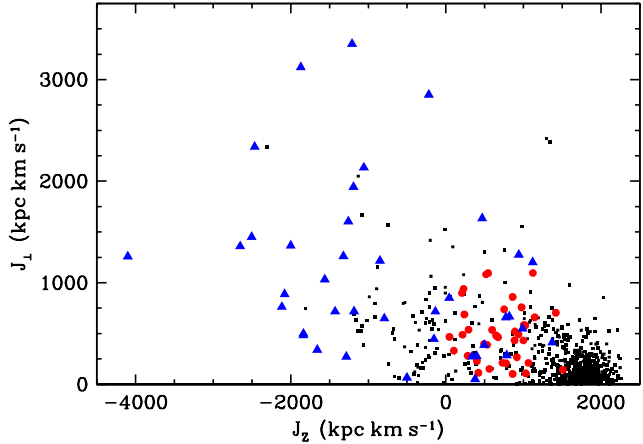}
\caption{
\label{angmomplot}
Inner and outer 
halo members as a function of their angular momentum components.
$J_{\perp}$ is defined as $(J_{X}^{2} + J_{Y}^{2})^{1/2}$.
Symbols are the same as in Figure~\ref{innerouterplot}.
[Please see the electronic edition for a color version of this figure.]
}
\end{figure}

In light of these arguments, it is perhaps even more remarkable
that, of the eight [X/Fe] ratios considered
for our halo populations in the previous sections, 
not a single abundance ratio of any inner halo star
deviates even minimally from 
the well-defined mean abundance trends of the inner halo population
(though [Na/Fe] shows an overall greater amount of scatter).
Even the heavy \ncap\ species obey a tight correlation with [Fe/H],
an abundance pattern rarely seen at [Fe/H]~$\lesssim -2.0$,
even in dSphs (e.g., \citealt{shetrone03}, A.\ Frebel et al., in prep.).
This is strong evidence for one of two possibilities.
Either the stars in our inner halo population
formed from a very homogenized ISM,
or the time-averaged abundance yields of various regions of the
stellar halo are nearly identical.
Both scenarios could also explain the apparent lack of very
low metallicity stars in the inner halo population, since 
it may have been common practice for stars with [Fe/H]~$\ll -2.0$ 
to form in regions of the halo where chemical enrichment was still
governed by local SNe events, rather than the time-averaged
yields of many SNe \citep[e.g.,][]{argast00}.

\subsection{Chemical Signatures of Possible Accretion Events}
\label{accretion}

\subsubsection{Blue Metal-Poor Stars}

Blue metal-poor (BMP) stars are the field analogs of the 
blue straggler stars found in clusters, 
except BMP's are believed to be formed by 
binary mass-transfer rather than stellar mergers 
\citep{preston00,sneden03b,carney05}.
\citet{preston94} have suggested that BMP stars 
may signify accretion events.
Based on new binary orbital solutions and analysis of their 
chemical compositions, however, \citet{preston00} and \citet{sneden03b} 
suggested that only the radial velocity (RV) constant BMP stars
may be intermediate-age stars that have been accreted, 
perhaps from satellite dSph systems.

Of the 175 stars examined by \citet{preston94}, two are present 
in our sample, \mbox{CS~22966--043} and \mbox{CS~22941--012}.
\mbox{CS~22966--043}, 
a long period binary with a nearly circular binary orbit 
($P = 317$~days, $e = 0.1$; \citealt{preston00}),
possesses a heretofore unique mix of the $\alpha$ elements
for a star with [Fe/H]~$=-1.91$: 
[Mg/Fe]~$=-0.65$, [Ca/Fe]~$=-0.24$, and [Ti~\textsc{ii}/Fe]~$= +0.49$
\citep{ivans03}.
\mbox{CS~22941--012}, a RV-constant star \citep{preston00}, 
was employed as a BMP, $\alpha$-normal 
comparison to \mbox{CS~22966--043} by \citet{ivans03}.
\mbox{CS~22966--043} barely did not pass our proper motion requirement.
If we relax this requirement and naively adopt a photometric distance,
\mbox{CS~22966--043} is on a very extreme orbit, only circling the 
Galactic center 4--8 times over 10 Gyr
($R_{\rm apo} = 152^{+65}_{-43}$~kpc and
$|Z_{\rm max}| = 75^{+26}_{-13}$~kpc).
\mbox{CS~22941--012}, in contrast, has orbital parameters 
$R_{\rm apo} = 26^{+3}_{-6}$~kpc and 
$|Z_{\rm max}| = 4.3\pm0.6$~kpc.
While the evolutionary origin of BMP binary stars is not in doubt,
it is interesting that \mbox{CS~22966--043}, a star with such 
a bizarre chemical composition, has an extreme orbit
that is consistent with the accretion scenario.
Meanwhile the RV-constant BMP star, \mbox{CS~22941--012},
appears to be on a more normal outer halo orbit.

\subsubsection{Dwarf Spheroidal Systems}

The number of dSph systems with known Galactic orbital
parameters is growing, thanks to careful measurements of their
proper motions.
The majority of the present-day dSphs with known proper motions and 
orbital parameters do not approach closer than 
a few tens of kpc from the Galactic center or the Solar neighborhood
(Canis Major: $R_{\rm peri} = 10.5\pm0.9$~kpc, \citealt{dinescu05};
Carina: $R_{\rm peri} = 20^{+43}_{-17}$~kpc, \citealt{piatek03};
Fornax: $R_{\rm peri} = 138\pm19$~kpc, \citealt{dinescu04};
$R_{\rm peri} = 118^{+19}_{-52}$~kpc, \citealt{piatek07};
Sculptor: $R_{\rm peri} = 120\pm51$~kpc, \citealt{dinescu04};
$R_{\rm peri} = 68^{+15}_{-37}$~kpc, \citealt{piatek06};
Ursa Minor: $R_{\rm peri} = 40^{+36}_{-30}$~kpc, \citealt{piatek05}).
Thus it is unlikely that these specific systems could have contributed
any significant fraction of the local stellar halo.
Additionally, the Leo~II dSph system has not undergone significant
tidal interaction with the Milky Way Galaxy, is currently at a very
great distance (218~kpc), and therefore might be on a more circular 
Galactic orbit that does not approach near to the Galactic center
\citep{siegel08}.
The orbital periods of the dSph systems are typically several Gyr, implying
that they will have reached their perigalactica no more than a few times,
and even the systems that do venture near the Solar radius would
only spend a small fraction of their orbital periods there.

Not coincidentally, this conclusion was also reached from 
comparison of the detailed abundances of several of these same dSphs 
(Carina, Fornax, Leo~II, Sculptor, and Ursa Minor) 
with halo stars of the same metallicity ranges \citep{venn04,shetrone08}.
One exception is the Sagittarius dSph, whose leading tidal
arm has deposited debris within a few kpc of the present
Solar neighborhood \citep{majewski03}.
The stars that Sagittarius is presently contributing to the halo
are different than the stars in its residual core
\citep{bellazzini06,chou07,siegel07},
signaling that dSph systems of the present day may not resemble the stars 
that they have already lost to the stellar halo of the Galaxy.

While the growing consensus points to few accreted dSph stars 
constituting the local metal-poor stellar halo,
chemical and kinematic analysis of
large samples of in situ halo stars will be needed to determine
if a significant fraction of the true outer halo (e.g., stars that
never approach nearer to the Galactic center than, say, 20~kpc) 
is comprised of the remnants of former or present dSph systems.
This is a daunting observational challenge, but the results from such
a study would be extremely interesting.

\section{Kinematics as a Function of Abundances}
\label{kfa}

\subsection{$\alpha$-poor Stars}
\label{alphapoor}

Most metal-poor stars exhibit [$\alpha$/Fe] ratios that are
0.3--0.4~dex above the Solar ratio 
\citep[e.g.,][]{wallerstein62,edvardsson93,mcwilliam95b}.
A handful of metal-poor stars, however, have [$\alpha$/Fe] ratios that
are significantly lower than the standard plateau,
and a few such stars have [$\alpha$/Fe] ratios
that are well below the Solar ratio 
\citep[e.g.,][]{carney97,ivans03}.  
Similar $\alpha$ deficiencies have been observed in 
several nearby dSph galaxies
\citep{bonifacio00a,shetrone01,shetrone03,tolstoy03,fulbright04b,
sadakane04,geisler05,sbordone07,koch08,shetrone08},
prompting speculation that the $\alpha$-poor stars may be 
signatures of past accretion of dSphs.
Furthermore, stars associated with the Sagittarius dSph, 
which is presently interacting with the Milky Way 
\citep{ibata94,majewski03}, are very dissimilar to field stars
and are decidedly underabundant in $\alpha$- and Fe-peak elements
(at least in the metallicity range covered, $-1.0<$~[Fe/H]~$<+0.0$;
\citealt{bonifacio00a,mcwilliam05,sbordone07}). 
Only a few $\alpha$-poor field stars are known, and for this
small subset of the halo the accretion hypothesis 
deserves additional scrutiny.
With the recent report of a handful of new $\alpha$-poor 
metal-poor stars \citep{barklem05} from the stellar 
content of the Hamburg-ESO Survey \citep{frebel06,christlieb08},
we are in a position to reevaluate the kinematic properties 
of these stars to search for clues of their origin.

In the top panel of Figure~\ref{alphapoorplot} 
we identify two classes of $\alpha$-poor stars: 
those with $+0.0 \leq$~[Mg/Fe]~$<+0.1$ (``$\alpha$-deficient'')
and those with [Mg/Fe]~$<+0.0$ (``$\alpha$-poor'').
For purposes of this analysis we also restrict our sample to 
stars with [Fe/H]~$<-1.0$.
In the bottom panels of Figure~\ref{alphapoorplot} we display the
kinematic and orbital properties of these stars.
They share no common region in phase space.
Several orbits extend to large distances above the Galactic plane 
or to large radii, but no kinematic signature is preferred.
This suggests that
at least a very small fraction of Galactic halo stars formed in
chemically inhomogeneous regions that were deficient (and,
allegedly, deficient by varying degrees) in the $\alpha$ elements.
\citet{ivans03} performed an exhaustive comparison of the abundance patterns 
of three $\alpha$-poor stars to Type~Ia SN models.
They concluded that
these stars seem to have increased contributions from Type~Ia yields
(relative to other stars at their metallicity), yet they cautioned that
no model could reproduce the overall abundance patterns of these stars.
They speculated that these stars may have been among the earliest to 
from from Type~Ia products, though they could also not rule out the
accretion hypothesis.

\begin{figure}
\epsscale{1.15}
\plotone{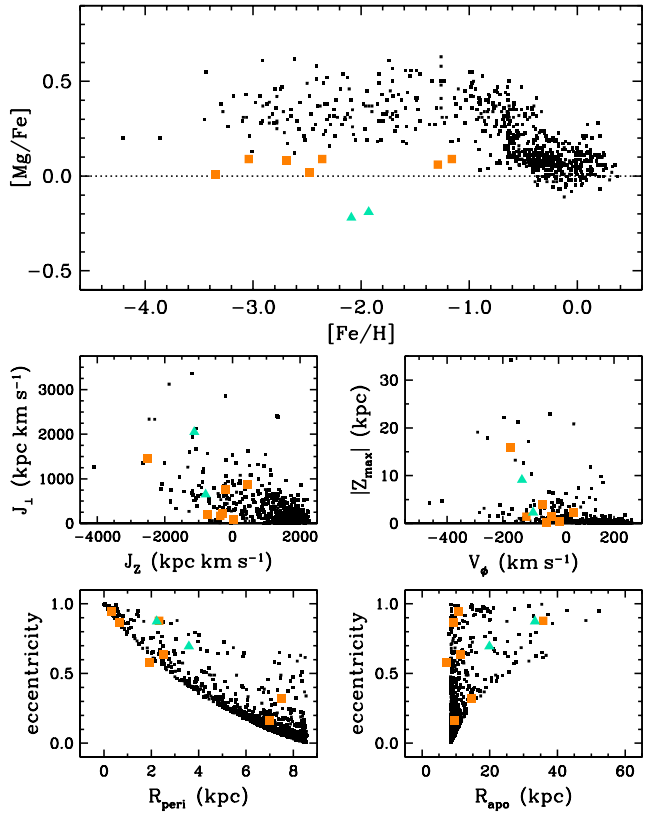}
\caption{
\label{alphapoorplot}
Defining stars of low [Mg/Fe].
Restricting our Mg-poor sample to stars with [Fe/H]~$< -1.0$, 
filled squares (orange in the online edition) represent stars with 
$+0.0 <$~[Mg/Fe]~$<+0.1$ and 
filled triangles (turquoise in the online edition) 
represent stars with [Mg/Fe]~$<+0.0$.
The stars selected according to these definitions are highlighted in the
lower four panels.
[Please see the electronic edition for a color version of this figure.]
}
\end{figure}

A number of studies
(\citealt{unavane96}, \citealt{fulbright02}, \citealt{venn04},
\citealt{pritzl05}, and \citealt{geisler07}, in addition to those listed above)
have concluded that the chemical compositions of dSph stars and halo giants 
are sufficiently different, and only a small fraction of the Milky Way stellar
halo can be composed of stars accreted from dSphs chemically
similar to those surviving to the present day.
Our result does not imply that the $\alpha$-poor stars could not
have formed in dSph systems that were assumed into the Milky Way.
Since many of these stars are on eccentric orbits that take them 
to large Galactic radii, we may surmise that either these inhomogeneous
regions were located at large Galactic radii (perhaps before the 
bulk of the Galaxy formed) or the stars were
formed in (separate?) dSphs and accreted at some time in the past.

\subsection{Stars with Specific \ncap\ Enrichment Signatures}
\label{rspro}

Can the \ncap\ material present in these metal-poor stars
be used to identify any preferred kinematics of the parent
clouds from which they formed?
Given the different mass ranges of the stars commonly thought
to produce \ncap\ material, this is an attractive possibility 
if the IMF's of the parent clouds differ significantly.

In the top panel of Figure~\ref{rsplot}, we identify stars
that have either a nearly pure \spro\ or \rpro\ [Ba/Eu] ratio.
Pure \rpro\ enrichment (or, viewed another way, lack of 
appreciable \spro\ enrichment) 
occurs over a very wide range of metallicity
($-3.0 <$~[Fe/H]~$-0.4$, or a factor of $\approx$~400 in Fe/H---a 
fact which is remarkable in its own right!).
Our sample includes a rather small number of stars exhibiting
a pure \spro\ signature; this is a consequence of our (somewhat
arbitrary) choice of literature data to include.
In the lower panels of Figure~\ref{rsplot}, we show the kinematic properties
of these stars.
The stars dominated by \rpro\ \ncap\ enrichment show no preferred
kinematic properties---they scatter over all values of 
$V_{\phi}$, $|Z_{\rm max}|$, $R_{\rm peri}$, and $R_{\rm apo}$.
The three stars exhibiting a pure \spro\ signature do not 
indicate any kinematic preferences, either.

\begin{figure}
\epsscale{1.15}
\plotone{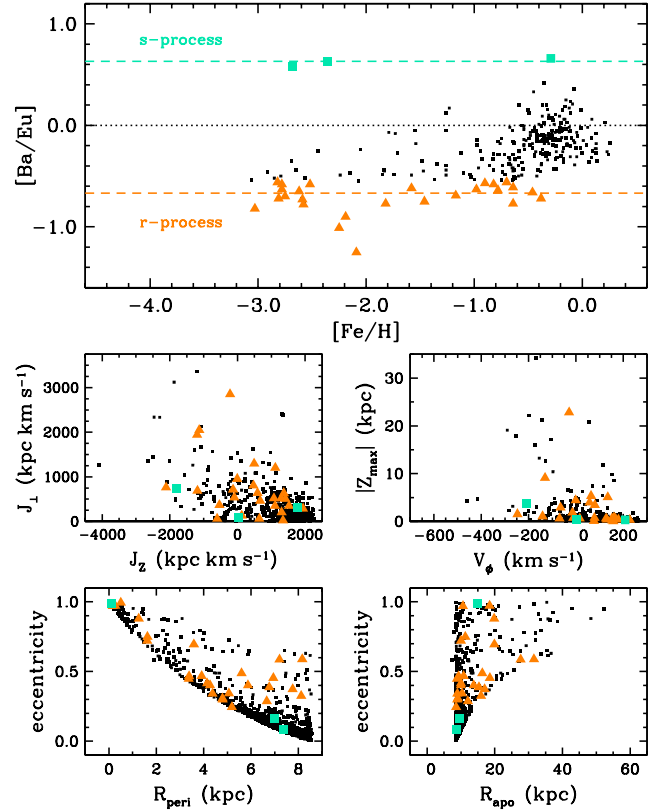}
\caption{
\label{rsplot}
Defining stars exhibiting pure \rpro\ and pure \spro\ 
[Ba/Eu] ratios.
The lower dashed line (orange in the online edition) 
represents the pure \rpro\ ratio ($-$0.67) and 
the upper dashed line (turquoise in the online edition)
represents the pure \spro\ ratio ($+$0.63)
as predicted by \citet{simmerer04}.
Filled triangles (orange in the online edition)
represent stars with [Ba/Eu]~$< -0.55$ and
filled squares (turquoise in the online edition) 
represent stars with [Ba/Eu]~$>+0.50$.
The stars selected according to these definitions are highlighted in the
lower four panels.
[Please see the electronic edition for a color version of this figure.]
}
\end{figure}

The main \rpro\ produces a robust abundance pattern
for species with $Z \geq 56$ 
\citep[e.g.,][]{sneden03a,ivans06,cowan06,roederer08}, 
but it has become clear over the
last decade that this unique signature does not apply to 
the lighter \ncap\ nuclei, such as the Sr-Y-Zr group.
Close analyses of two metal-poor stars, \mbox{HD~88609}
\citep{honda07} and \mbox{HD~122563} \citep{honda06}, reveal
an abundance pattern that gradually declines with increasing $Z$,
producing high ratios between light and heavy \ncap\ species
(see also \citealt{mcwilliam95b}, \citealt{wasserburg96},
\citealt{johnson02}, \citealt{aoki05},
\citealt{lai07}, and \citealt{lai08}).
This pattern cannot be associated with the main \rpro,
the main \spro, or the weak \spro, and could be representative of---for
example---a light element primary process \citep{travaglio04},
a rapid proton-capture process \citep{wanajo06a},
a weak \rpro\ \citep{wanajo06b},
a cold \rpro\ \citep{wanajo07},
hypernovae \citep{qian08}, or 
a high entropy wind from Type~II SNe \citep{farouqi08}.
This multiplicity of scenarios, with a variety of proposed 
SN mass ranges responsible for the creation of the
light \ncap\ elements, demands additional observational constraints.
Once again, if the IMF of a certain proto-stellar metal-poor
cloud was significantly different from other clouds, this could
manifest itself in the derived [Ba/Y] ratios.
In the top panel of Figure~\ref{weakrplot} we identify stars
with low [Ba/Y] ratios ([Ba/Y]~$<-0.55$) and low [Ba/Fe] ratios
([Ba/Fe]~$<-0.50$), similar to the abundances found in 
\mbox{HD~88609} and \mbox{HD~122563}.\footnote{ 
We note, as \citet{honda07} have, that high
[Sr/Ba] or [Y/Ba] ratios do not uniquely define the abundance pattern 
associated with this process, since \mbox{HD~88609} was
selected for analysis based on its [Sr/Ba] ratio, 
identical to that of \mbox{HD~122563}.}
Again, though, these stars exhibit no clear kinematic properties,
although they may slightly prefer eccentric orbits, which 
could point to a common---though yet poorly constrained---origin.
(The common retrograde orbits are a result of the narrow
[Fe/H] range of this sub-sample, and are not of physical consequence.)

\begin{figure}
\epsscale{1.15}
\plotone{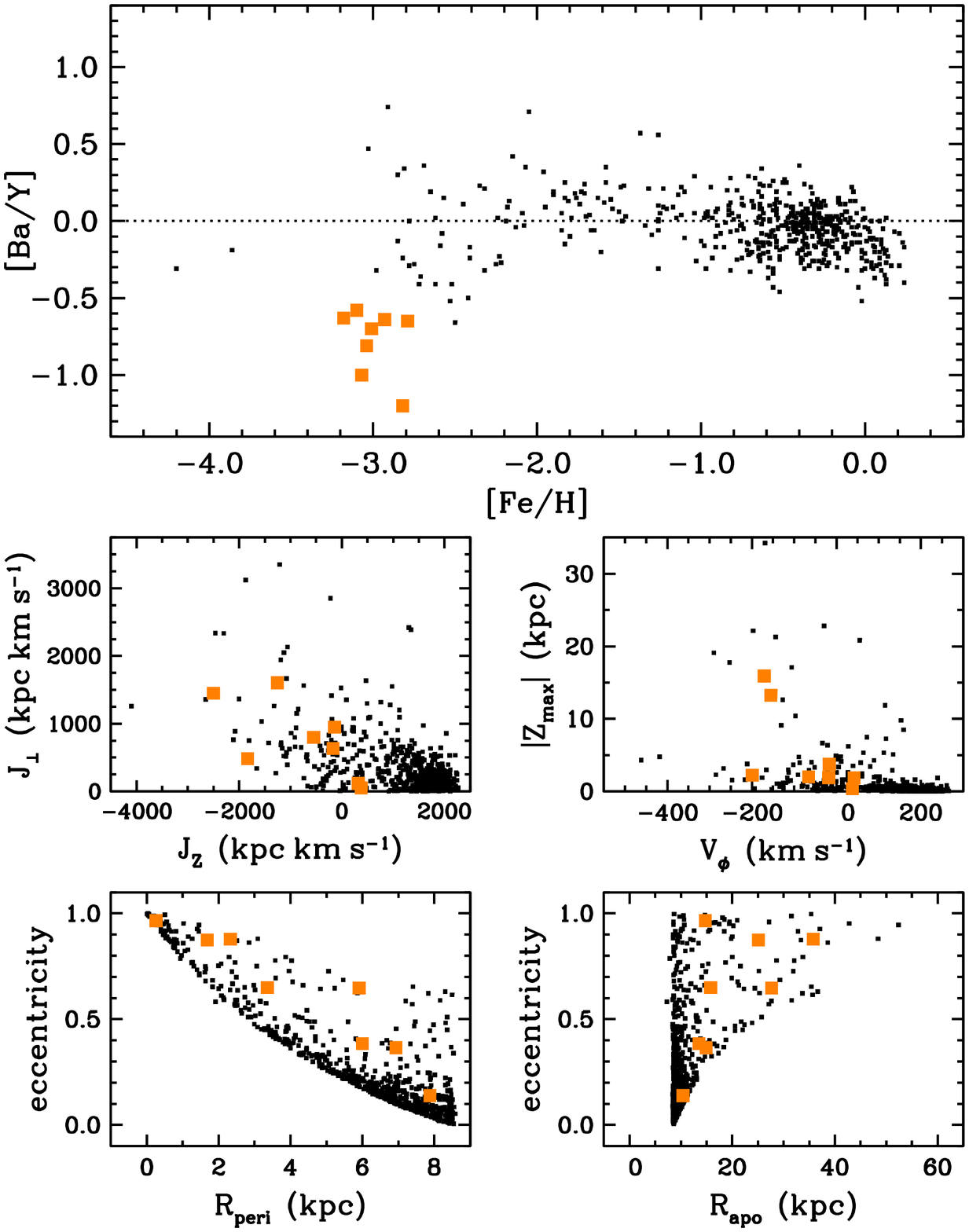}
\caption{
\label{weakrplot}
Defining stars with excesses of the light \ncap\ elements and
deficiencies of the heavy \ncap\ elements.
The filled squares (orange in the online edition)
represent stars with [Ba/Y]~$<-0.55$ and [Ba/Fe]~$<-0.50$.
(The [Ba/Y] ratio is $-$0.69 and $-$0.68 in \mbox{HD~88609} and 
\mbox{HD~122563}, respectively; \citealt{honda06,honda07}.)
The stars selected according to this definition are highlighted in the
lower four panels.
[Please see the electronic edition for a color version of this figure.]
}
\end{figure}

At low metallicities, \ncap\ enrichment is probably a very
localized phenomenon that results in a wide distribution of
\ncap\ abundances, and thus it will be extremely difficult
to identify any associated large-scale kinematic behaviors
of the proto-stellar clouds from which these stars formed.
The potential gain from such a connection, however, justifies
the continued effort toward this goal.

\subsection{The Most Metal-Poor Stars Known}
\label{ultrampstars}

We also examine the kinematic properties of the three most
metal-poor stars known, 
\mbox{HE~0107--5240} ([Fe/H]~$=-5.3$; \citealt{christlieb02,christlieb04}),
\mbox{HE~1327--2326} ([Fe/H]~$=-5.96$; \citealt{frebel05,aoki06,frebel08a}), 
and
\mbox{HE~0577--4840} ([Fe/H]~$=-4.75$; \citealt{norris07}).
Proper motions have been reported for all three of these stars
in the UCAC2 or NOMAD catalogs; unfortunately, for 
\mbox{HE~0107--5240} and \mbox{HE~0577--4840} the measurements
are no larger than their uncertainties ($\sim 6$~mas~yr$^{-1}$) and are
therefore not useful for our purposes.
The proper motions for \mbox{HE~1327--2326} are large:
$\mu_{\alpha} = -40\pm1$, $\mu_{\delta} = 58\pm4$~mas~yr$^{-1}$
\citep{zacharias04b}.
We compute a photometric distance of 1160~pc for \mbox{HE~1327--2326}
(assuming it is a subgiant and not a dwarf; see 
\citealt{korn08} and \citealt{frebel08a}), 
in fair agreement with the previous estimate of 1400~pc \citep{aoki06}.
\mbox{HE~1327--2326} lies above the plane of the Galaxy
($\ell = 314^{\circ}$, $b = +38^{\circ}$) and is moving rapidly
away from the Galactic plane ($U = 221$, $V = 246$, $W=325$, all in \kmsec).
Using these input parameters, \mbox{HE~1327--2326} is only weakly
bound to the gravitational potential of the Galaxy, completing no more than
2~orbits over 10~Gyr.  
Altering the distance by $\pm$500~pc (much larger than the 20\% uncertainty
assumed in \S~\ref{uncertainties}),
proper motion by $\pm$5~mas~yr$^{-1}$, 
radial velocity by $\pm$10~\kmsec, and 
mass of the Galaxy by $\pm$10\%, we find that this star
completes between 1 and 4~orbits over 10~Gyr.
For these orbits, $R_{\rm peri} = 8.0\pm0.2$~kpc,
$R_{\rm apo} = 405^{+265}_{-186}$~kpc, and
$|Z_{\rm max}| = 160^{+97}_{-45}$.

All reasonable variations in the input parameters still point to 
a very extreme orbit for \mbox{HE~1327--2326}, which is perhaps 
not unexpected given the uncharacteristically low metal abundance
of this interesting star.
Precise proper motion measurements for \mbox{HE~0107--5240} and
\mbox{HE~0577--4840} will be necessary to determine whether extreme
orbits are a common characteristic of the most metal-poor stars in 
the Galaxy.
Furthermore, \mbox{HE~1327--2326} spends a very small fraction of
its time in the inner regions of the Galaxy, suggesting, perhaps, that
many more objects with compositions similar to this star
currently reside in the distant realms of the stellar halo.
This again hints to the possibility that
objects in the true outer halo of the Galaxy may bear
little resemblance to those that inhabit the inner few tens of kpc.

\section{Conclusions}
\label{conclusions}

We have compiled a large sample of metal-poor 
stars with reported abundances from the literature.
For the subset of these stars with reliable proper motion measurements,
we have computed space velocities and Galactic orbital parameters.
We have used the combination of chemical and kinematic information 
to identify any abundance patterns that are common to stars with 
similar kinematics or identify any kinematic signatures that are common
to stars with similar chemical enrichment patterns. 
Our major results can be summarized as follows.

(1) A proper motion selection criterion for our sample 
biases the very metal-poor end of our sample toward stars 
with large proper motions and hence extreme orbits, 
many of which are on retrograde Galactic orbits.
This bias must be borne in mind when interpreting chemical
enrichment patterns in very metal-poor stars.

(2) We find no abundance trends with maximum radial distance from
the Galactic center or maximum vertical distance above the Galactic disk.
Current high-resolution abundance analyses are limited to sampling
halo stars that happen to be passing relatively near to the 
Solar neighborhood.
Other studies have found very slight decreases in abundance ratios
with increasing distance from the Galactic center, and this issue may
remain unsettled until large, deep abundance surveys and proper motion
surveys can reach beyond the 
Solar neighborhood into a true in situ sample of the stellar halo.

(3) We use only kinematic criteria to define our inner halo
(stars on prograde, eccentric orbits that do not stray beyond 
15~kpc from the Galactic center or 5~kpc from the Galactic plane)
and outer halo (stars on very retrograde orbits or stars whose
orbits reach more than 25~kpc from the Galactic center or
more than 10~kpc above the Galactic plane) samples, based
on the kinematic properties of these populations derived from
much larger datasets.
In the metallicity regimes where the two populations overlap,
roughly $-2.5 <$~[Fe/H]~$<-1.5$,
the [Mg/Fe] ratio of the outer halo may be lower than 
the inner halo by $\sim$~0.1~dex.
For [Ni/Fe] and [Ba/Fe]
the star-to-star abundance scatter of the inner halo is consistently
smaller than the star-to-star abundance scatter of the outer halo.
The [Na/Fe], [Ca/Fe], [Ti/Fe], and [Y/Fe] 
ratios of both populations show similar levels of scatter.
We do not have enough [Eu/Fe] measurements in our sample to draw
any conclusions from this abundance ratio.

(4) Our inner halo population appears chemically homogeneous, 
suggesting that a significant fraction
of the Milky Way stellar halo had a common origin from a 
well-mixed ISM.

(5) In contrast, our kinematically diverse outer halo population is 
also chemically diverse, suggesting that another significant fraction
of the Milky Way stellar halo formed in remote regions where
chemical enrichment was dominated by local SN events.
This component is reminiscent of the ``chaotic origin'' for the
Galaxy suggested by the globular cluster data of \citet{searle78}.

(6) If we classify globular clusters by these same kinematic 
criteria, all of the inner halo and most of the outer halo clusters 
follow similar (mean) abundance trends with comparable degrees of scatter 
to the inner halo population of field stars.
The chemical similarity of the inner halo and the globular clusters 
may suggest that
these abundance trends represent the time-averaged mean abundances
of a metal-poor ISM pre-enriched by the earliest generations of stars.
We find no mean cluster abundance dependence with increasing radial
distance from the Galactic center or vertically from the Galactic plane.

(7) We find no kinematic signature in common to groups of metal-poor
stars with peculiar abundance patters ($\alpha$-poor stars,
stars showing a pure-$s$- or pure-$r$-process \ncap\ enrichment pattern,
or stars with a deficiency of heavy \ncap\ material).

(8) Our results do not exclude the possibility that any of these 
individual stars
were accreted by the Milky Way from, e.g., dSph systems;
however, the orbits of many present-day dSphs (especially those 
whose stars have been subject to high-resolution abundance analyses) 
rarely bring them near the Solar radius, 
so they would not be expected to contribute
many stars to the local metal-poor stellar halo.

(9) Several individual stars---including the most metal-poor star 
known---and dSph systems whose compositions differ greatly from 
the bulk of the stellar halo in the Solar neighborhood
have long orbital periods ($\sim$ few Gyr) 
and extreme orbital characteristics ($R_{\rm apo} \sim$ few hundred kpc).
If stars like these spend the majority of their time in the distant
regions of the Milky Way stellar halo, this raises the possibility that
many more stars with unusual abundance patterns may occupy the 
true outer halo of the Galaxy, which may have little resemblance
to the local stellar halo.

The abundance dataset used for this analysis was compiled from a variety 
of literature sources, and the inherent systematic differences
from one study to another limit our ability to detect more subtle
chemical differences than those described above.
Large, metal-poor stellar samples from which abundances are derived in a 
homogeneous manner are necessary to perform a more detailed 
nucleo-kinematic analysis.
The construction and analysis of such datasets is presently underway.

Evidence continues to grow in support of the notion 
that the chemical enrichment history 
of the Milky Way stellar halo is nonuniform on both small and large scales.
While it is unlikely that an exact correlation
between the kinematic properties of a star and its chemical abundance
pattern will ever be identified, 
we may be approaching an era where it is necessary to 
know the kinematic properties of a field halo star in order to 
place its chemical abundance pattern in the proper nucleosynthetic context.
The challenge for future studies will be to articulate the degree 
of chemical dissimilarity in the Milky Way halo kinematic substructure.

\acknowledgments

It is a pleasure to thank Anna Frebel and Chris Sneden 
for many extensive discussions relating to this topic and 
detailed comments on assorted versions of this manuscript.
I.~U.~R.\ also thanks John Cowan, George Preston, and Matthew Shetrone 
for their helpful suggestions and encouragement, as well as
David Lambert, who originally suggested this literature study.
Additional thanks go to Monique Spite for sending radial velocity
measurements for the ``First Stars'' sample in advance of publication,
Douglas Lin for sharing a copy his orbit integrator code, and 
the anonymous referee for helpful suggestions.
This research would not have been possible without the resources available
from the SIMBAD and VizieR \citep{ochsenbein00} online databases, 
operated at CDS, Strasbourg, France, 
the Two Micron All-Sky Survey, 
and NASA's Astrophysics Data System Bibliographic Services.
Funding has been provided by NSF grant AST~06-07708 (to C.~Sneden).

\clearpage

\begin{deluxetable}{lrrrrrrrrrrrrr} 
\tablecaption{Space Velocities and Orbital Parameters for the Stellar Sample
\label{startab}}
\tablewidth{0pt}
\tabletypesize{\scriptsize}
\tablehead{
\colhead{Star name} &
\colhead{[Fe/H]} &
\colhead{$D$} &
\colhead{$U$} &
\colhead{$V$} &
\colhead{$W$} &
\colhead{$V_{\phi}$} &
\colhead{$R_{\rm peri}$} &
\colhead{$R_{\rm apo}$} &
\colhead{$e$} &
\colhead{$|Z_{\rm max}|$} &
\colhead{$J_{X}$} &
\colhead{$J_{Y}$} &
\colhead{$J_{Z}$} \\
\colhead{} &
\colhead{} &
\colhead{(pc)} &
\colhead{(\kmsec)} &
\colhead{(\kmsec)} &
\colhead{(\kmsec)} &
\colhead{(\kmsec)} &
\colhead{(kpc)} &
\colhead{(kpc)} &
\colhead{} &
\colhead{(kpc)} &
\colhead{(kpc \kmsec)} &
\colhead{(kpc \kmsec)} &
\colhead{(kpc \kmsec)}}
\startdata
       BD~$+$17~3248 & $-$2.05 &   243 &   75 &   99 &  $-$21 &   98 &  7.9 & 20.6 & 0.45 &   0.3 &   $-$13 &   181 &   817 \\ 
        CD~$-$38~245 & $-$4.20 &  4728 &  261 & $-$129 &  $-$18 & $-$106 &  2.5 & 36.3 & 0.87 &  17.1 &  $-$587 & $-$1065 &  $-$850 \\ 
      CS~31082$-$001 & $-$2.91 &  2515 & $-$150 & $-$256 & $-$206 & $-$253 &  1.2 & 20.0 & 0.89 &  17.8 &  $-$661 &  2239 & $-$2301 \\ 
            G~4$-$36 & $-$1.93 &   244 &  337 &  $-$89 &  $-$81 &  $-$91 &  2.2 & 33.4 & 0.87 &   2.3 &   $-$19 &   648 &  $-$793 \\ 
         HD~122563 & $-$2.79 &   237 &  112 &   38 &   17 &   38 &  6.9 & 14.9 & 0.37 &   0.3 &    $-$8 &  $-$117 &   320 \\ 
      HE~1327$-$2326 & $-$5.96 &  1160 &  221 &  246 &  325 &  263 & \nodata & \nodata & \nodata & \nodata &  $-$389 & $-$2398 &  2076 \\ 
         HIP~92167 & $-$1.47 &   162 &   61 &  152 & $-$284 &  153 &  7.1 & 13.5 & 0.31 &   9.8 &   $-$51 &  2421 &  1307 \\ 
           HR~2835 & $-$0.55 &    31 &   62 &  217 &  $-$26 &  217 &  6.9 & 10.2 & 0.19 &   0.2 &    $-$2 &   218 &  1848 \\ 
\enddata
\tablecomments{
The full table is available in machine-readable form in the electronic
edition of the journal;
only a small portion is shown here to present its general form and content.
Only stars that made our proper motion cut are shown in the table.
No orbital parameters are given for stars that did not complete at least
20 orbits of the Galactic center.}
\end{deluxetable}

\clearpage
\begin{deluxetable}{lcccccccccc} 
\tablecaption{Abundance Ratios for Members of the Inner Halo Population
\label{innerabundtab}}
\tablewidth{0pt}
\tabletypesize{\scriptsize}
\tablehead{
\colhead{Star name} &
\colhead{[Fe/H]}  &
\colhead{[Na/Fe]} &
\colhead{[Mg/Fe]} &
\colhead{[Ca/Fe]} &
\colhead{[Ti/Fe]} &
\colhead{[Ni/Fe]} &
\colhead{[Y/Fe]}  &
\colhead{[Ba/Fe]} &
\colhead{[Eu/Fe]} &
\colhead{Reference}} 
\startdata
HD~221830     &  $-$0.52 &  $+$0.14 &  $+$0.47 &  $+$0.24 &  $+$0.25 &  $+$0.15 &  $-$0.03 &  $-$0.28 &  \nodata  &   1 \\
HIP~96185     &  $-$0.58 &  $+$0.19 &  $+$0.38 &  $+$0.21 &  $+$0.32 &  $+$0.06 &  $-$0.06 &  $+$0.04 &  $+$0.29  &   2 \\
HD~157089     &  $-$0.59 &  $+$0.04 &  $+$0.22 &  $+$0.18 &  $+$0.27 &  $+$0.00 &  $+$0.02 &  $-$0.28 &  \nodata  &   1 \\
HIP~86431     &  $-$0.64 &  $+$0.08 &  $+$0.33 &  $+$0.17 &  $+$0.20 &  $+$0.01 &  $-$0.15 &  $-$0.07 &  $+$0.22  &   2 \\
HIP~58357     &  $-$0.65 &  $+$0.04 &  $+$0.31 &  $+$0.18 &  $+$0.32 &  $+$0.03 &  \nodata &  $+$0.16 &  \nodata  &   2 \\
HIP~33582     &  $-$0.74 &  $+$0.06 &  $+$0.47 &  $+$0.24 &  $+$0.41 &  \nodata &  \nodata &  $-$0.18 &  \nodata  &   2 \\
HIP~64115     &  $-$0.74 &  $+$0.15 &  $+$0.47 &  $+$0.27 &  $+$0.29 &  $+$0.09 &  $+$0.04 &  $-$0.01 &  $+$0.39  &   2 \\
HIP~92781     &  $-$0.75 &  $+$0.09 &  $+$0.37 &  $+$0.26 &  $+$0.25 &  $+$0.04 &  $+$0.09 &  $+$0.24 &  $+$0.37  &   2 \\
HIP~74033     &  $-$0.78 &  $+$0.13 &  $+$0.34 &  $+$0.21 &  $+$0.29 &  $+$0.00 &  $+$0.07 &  $+$0.01 &  $+$0.30  &   2 \\
HIP~58962     &  $-$0.80 &  $-$0.04 &  $+$0.11 &  $+$0.06 &  $+$0.01 &  $-$0.07 &  $-$0.19 &  $-$0.18 &  \nodata  &   3 \\
HIP~86013     &  $-$0.82 &  $+$0.14 &  $+$0.39 &  $+$0.28 &  $+$0.34 &  $+$0.04 &  $+$0.00 &  $-$0.07 &  $+$0.37  &   2 \\
HIP~58229     &  $-$0.94 &  $-$0.14 &  $+$0.18 &  $+$0.20 &  $+$0.20 &  $-$0.10 &  $-$0.15 &  $+$0.07 &  $+$0.32  &   2 \\
HIP~66665     &  $-$0.97 &  $+$0.16 &  $+$0.49 &  $+$0.34 &  $+$0.25 &  $+$0.08 &  $-$0.10 &  $-$0.06 &  $+$0.08  &   2 \\
HIP~77946     &  $-$0.97 &  $+$0.21 &  $+$0.55 &  $+$0.36 &  $+$0.35 &  $+$0.09 &  $+$0.00 &  $+$0.05 &  $+$0.24  &   2 \\
HIP~109067    &  $-$0.97 &  $+$0.03 &  $+$0.36 &  $+$0.32 &  $+$0.25 &  $+$0.07 &  $+$0.21 &  $-$0.10 &  \nodata  &   2 \\
HIP~10449     &  $-$0.98 &  $-$0.04 &  $+$0.42 &  $+$0.28 &  $+$0.30 &  $+$0.01 &  $+$0.08 &  $+$0.10 &  \nodata  &   2 \\
HIP~5458      &  $-$1.04 &  $-$0.06 &  $+$0.45 &  $+$0.37 &  $+$0.12 &  $+$0.02 &  $-$0.21 &  $+$0.08 &  $+$0.56  &   2 \\
HIP~17666     &  $-$1.10 &  $-$0.05 &  $+$0.46 &  $+$0.29 &  $+$0.36 &  $+$0.05 &  $+$0.55 &  $+$0.44 &  \nodata  &   2 \\
HIP~44033     &  $-$1.12 &  $-$0.06 &  $+$0.32 &  $+$0.23 &  $+$0.22 &  $-$0.07 &  $-$0.05 &  $+$0.17 &  \nodata  &   4 \\
HIP~104659    &  $-$1.12 &  $+$0.08 &  $+$0.38 &  $+$0.21 &  $+$0.27 &  $+$0.01 &  $-$0.07 &  $-$0.03 &  $+$0.24  &   2 \\
HIP~98532     &  $-$1.23 &  $-$0.12 &  $+$0.50 &  $+$0.37 &  $+$0.31 &  $+$0.07 &  $+$0.14 &  $+$0.23 &  $+$0.06  &   2 \\
HIP~81170     &  $-$1.26 &  $-$0.11 &  $+$0.41 &  $+$0.27 &  $+$0.39 &  $+$0.00 &  $+$0.37 &  $+$0.06 &  $+$0.50  &   2 \\
HIP~109390    &  $-$1.34 &  $-$0.05 &  $+$0.50 &  $+$0.23 &  $+$0.29 &  $-$0.08 &  $+$0.04 &  $+$0.01 &  $+$0.39  &   2 \\
HIP~60632     &  $-$1.65 &  $+$0.04 &  $+$0.43 &  $+$0.36 &  $+$0.44 &  $+$0.07 &  $+$0.08 &  $+$0.02 &  \nodata  &   2 \\
HIP~98020     &  $-$1.67 &  $-$0.29 &  $+$0.26 &  $+$0.27 &  $+$0.28 &  $+$0.00 &  $-$0.05 &  $-$0.11 &  \nodata  &   2 \\
HIP~29992     &  $-$1.71 &  $-$0.08 &  $+$0.33 &  $+$0.28 &  $+$0.25 &  $-$0.03 &  \nodata &  $-$0.27 &  \nodata  &   2 \\
HIP~97468     &  $-$1.71 &  $+$0.01 &  $+$0.57 &  $+$0.44 &  $+$0.35 &  $+$0.01 &  $-$0.18 &  $+$0.00 &  $+$0.27  &   2 \\
HIP~38621     &  $-$1.81 &  $-$0.07 &  $+$0.43 &  $+$0.40 &  $+$0.27 &  $+$0.04 &  \nodata &  $-$0.33 &  \nodata  &   2 \\
HIP~103269    &  $-$1.81 &  $-$0.41 &  $+$0.29 &  $+$0.23 &  $+$0.24 &  $-$0.04 &  $-$0.13 &  $+$0.00 &  \nodata  &   2 \\
HIP~68807     &  $-$1.83 &  $-$0.13 &  $+$0.49 &  $+$0.37 &  $+$0.20 &  $-$0.03 &  $-$0.30 &  $-$0.05 &  $+$0.40  &   2 \\
HIP~18915     &  $-$1.85 &  $+$0.14 &  $+$0.38 &  $+$0.36 &  $+$0.35 &  $+$0.07 &  $+$0.17 &  $+$0.11 &  \nodata  &   2 \\
HIP~44124     &  $-$1.96 &  \nodata &  $+$0.37 &  $+$0.24 &  $+$0.25 &  $+$0.12 &  \nodata &  $+$0.10 &  \nodata  &   2 \\
HIP~14594     &  $-$2.13 &  $-$0.05 &  $+$0.52 &  $+$0.35 &  $+$0.37 &  $+$0.05 &  $-$0.10 &  $-$0.14 &  \nodata  &   2 \\
HIP~115949    &  $-$2.19 &  $-$0.50 &  $+$0.46 &  $+$0.38 &  $+$0.25 &  $+$0.10 &  $-$0.28 &  $-$0.28 &  $+$0.62  &   2 \\
HIP~60719     &  $-$2.35 &  $+$0.11 &  $+$0.49 &  $+$0.39 &  $+$0.24 &  \nodata &  $-$0.25 &  $-$0.02 &  \nodata  &   2 \\
HIP~96115     &  $-$2.41 &  $+$0.08 &  $+$0.58 &  $+$0.51 &  $+$0.49 &  $+$0.23 &  $+$0.08 &  $-$0.16 &  \nodata  &   2 \\
HIP~72461     &  $-$2.48 &  $-$0.07 &  $+$0.42 &  $+$0.37 &  $+$0.52 &  \nodata &  \nodata &  $-$0.32 &  \nodata  &   2 \\
\enddata
\tablerefs{
(1)~\citealt{edvardsson93};
(2)~\citealt{fulbright00};
(3)~\citealt{nissen97};
(4)~\citealt{stephens02}
}
\end{deluxetable}

\clearpage
\begin{deluxetable}{lcccccccccc} 
\tablecaption{Abundance Ratios for Members of the Outer Halo Population
\label{outerabundtab}}
\tablewidth{0pt}
\tabletypesize{\scriptsize}
\tablehead{
\colhead{Star name} &
\colhead{[Fe/H]}  &
\colhead{[Na/Fe]} &
\colhead{[Mg/Fe]} &
\colhead{[Ca/Fe]} &
\colhead{[Ti/Fe]} &
\colhead{[Ni/Fe]} &
\colhead{[Y/Fe]}  &
\colhead{[Ba/Fe]} &
\colhead{[Eu/Fe]} &
\colhead{Reference}} 
\startdata
HIP~19814         &  $-$0.71 &  $-$0.48 &  $+$0.07 &  $+$0.01 &  $+$0.08 &  $-$0.13 &  $-$0.12 &  $+$0.17 &  \nodata &    1 \\
HIP~117041        &  $-$0.88 &  $+$0.12 &  $+$0.53 &  $+$0.32 &  $+$0.30 &  $+$0.09 &  $-$0.02 &  $-$0.10 &  \nodata &    2 \\
HIP~57265         &  $-$1.10 &  $-$0.29 &  $+$0.19 &  $+$0.26 &  $+$0.17 &  $-$0.10 &  $-$0.18 &  $+$0.01 &  $+$0.48 &    2 \\
HIP~62747         &  $-$1.54 &  $+$0.00 &  $+$0.56 &  $+$0.40 &  $+$0.26 &  $+$0.02 &  $+$0.00 &  $+$0.08 &  $+$0.26 &    2 \\
HD~20		  &  $-$1.58 &  \nodata &  $+$0.17 &  $+$0.24 &  $+$0.20 &  $-$0.09 &  $-$0.04 &  $+$0.31 &  $+$0.80 &    3 \\
HIP~5445          &  $-$1.58 &  $-$0.33 &  $+$0.26 &  $+$0.27 &  $+$0.24 &  $-$0.09 &  $-$0.22 &  $+$0.03 &  $+$0.65 &    2 \\
HIP~19797         &  $-$1.68 &  $+$0.09 &  $+$0.34 &  $+$0.32 &  $+$0.38 &  $+$0.00 &  $+$0.00 &  $+$0.04 &  \nodata &    2 \\
HE~0447$-$4858	  &  $-$1.69 &  \nodata &  $+$0.24 &  $+$0.24 &  $+$0.28 &  $+$0.33 &  \nodata &  $+$0.45 &  \nodata &    3 \\
HIP~15904         &  $-$1.76 &  $-$0.14 &  $+$0.40 &  $+$0.33 &  $+$0.30 &  $+$0.01 &  $-$0.14 &  $+$0.01 &  \nodata &    1 \\
HIP~21609         &  $-$1.76 &  $-$0.25 &  $+$0.34 &  $+$0.39 &  $+$0.63 &  $-$0.05 &  \nodata &  $-$0.30 &  \nodata &    2 \\
HIP~115167        &  $-$1.77 &  $+$0.10 &  $+$0.33 &  $+$0.43 &  $+$0.38 &  $+$0.09 &  $-$0.05 &  $+$0.13 &  \nodata &    2 \\
HE~1343$-$0640	  &  $-$1.90 &  \nodata &  $+$0.37 &  $+$0.41 &  $+$0.32 &  $-$0.21 &  $+$0.51 &  $+$0.70 &  \nodata &    3 \\
G~4$-$36          &  $-$1.93 &  $-$0.28 &  $-$0.19 &  $-$0.11 &  $+$0.55 &  $+$0.48 &  \nodata &  $-$0.72 &  \nodata &    4 \\
CS~29522$-$046    &  $-$2.09 &  \nodata &  $+$0.40 &  $+$0.38 &  $+$0.41 &  $-$0.03 &  $+$0.09 &  $+$0.14 &  \nodata &    5 \\
HIP~87693         &  $-$2.11 &  $+$0.12 &  $+$0.46 &  $+$0.39 &  $+$0.48 &  $+$0.06 &  \nodata &  $-$0.46 &  \nodata &    2 \\
HE~0143$-$1135	  &  $-$2.13 &  \nodata &  $+$0.33 &  $+$0.26 &  $+$0.24 &  $-$0.09 &  \nodata &  $-$0.07 &  \nodata &    3 \\
HD~221170	  &  $-$2.15 &  \nodata &  $+$0.30 &  $+$0.27 &  $+$0.24 &  $-$0.16 &  $-$0.04 &  $+$0.38 &  $+$0.85 &    3 \\
HE~2329$-$3702	  &  $-$2.15 &  \nodata &  $+$0.31 &  $+$0.44 &  $+$0.42 &  $-$0.22 &  \nodata &  $-$0.14 &  \nodata &    3 \\
HE~1330$-$0354	  &  $-$2.29 &  \nodata &  $+$0.32 &  $+$0.40 &  $+$0.54 &  $-$0.08 &  \nodata &  $-$0.47 &  \nodata &    3 \\
HIP~86443         &  $-$2.32 &  \nodata &  $+$0.52 &  $+$0.38 &  $+$0.42 &  $+$0.07 &  $+$0.07 &  $-$0.25 &  \nodata &    2 \\
HE~1337$-$0453	  &  $-$2.34 &  \nodata &  $+$0.38 &  $+$0.44 &  $+$0.39 &  $-$0.14 &  \nodata &  $-$0.24 &  \nodata &    3 \\
HE~1256$-$0651	  &  $-$2.35 &  \nodata &  $+$0.22 &  $+$0.32 &  $+$0.30 &  $+$0.06 &  \nodata &  $-$0.26 &  \nodata &    3 \\
HE~2345$-$1919	  &  $-$2.46 &  \nodata &  $+$0.33 &  $+$0.25 &  $+$0.22 &  $-$0.01 &  \nodata &  $-$0.45 &  \nodata &    3 \\
BD$+$24~1676      &  $-$2.50 &  $-$0.20 &  $+$0.49 &  $+$0.43 &  $+$0.39 &  $+$0.30 &  $+$0.17 &  $-$0.49 &  \nodata &    4 \\
HD~186478      	  &  $-$2.57 &  $+$0.20 &  $+$0.39 &  $+$0.44 &  $+$0.27 &  $-$0.18 &  $-$0.19 &  $-$0.04 &  $+$0.48 &    6 \\
HE~1320$-$1339	  &  $-$2.78 &  \nodata &  $+$0.25 &  $+$0.26 &  $+$0.25 &  $+$0.43 &  $-$0.13 &  $-$0.42 &  $+$0.16 &    3 \\
CS~29491$-$069    &  $-$2.81 &  \nodata &  $+$0.28 &  $+$0.29 &  $+$0.20 &  $-$0.05 &  $+$0.00 &  $+$0.34 &  $+$1.06 &    3 \\
CS~29495$-$041    &  $-$2.82 &  $+$0.24 &  $+$0.33 &  $+$0.38 &  $+$0.24 &  $+$0.02 &  $-$0.41 &  $-$0.65 &  $-$0.09 &    6 \\
BS~16085$-$050    &  $-$2.91 &  \nodata &  $+$0.61 &  $+$0.38 &  $+$0.22 &  $+$0.20 &  \nodata &  $-$1.56 &  \nodata &    7 \\
HE~0547$-$4539	  &  $-$3.01 &  \nodata &  $+$0.13 &  $+$0.13 &  $+$0.19 &  $+$0.03 &  $-$0.33 &  $-$1.03 &  \nodata &    3 \\
CS~22169$-$035    &  $-$3.04 &  \nodata &  $+$0.09 &  $+$0.13 &  $-$0.08 &  $-$0.28 &  $-$0.38 &  $-$1.19 &  \nodata &    6 \\
BD$-$18~5550      &  $-$3.06 &  $+$0.05 &  $+$0.31 &  $+$0.41 &  $+$0.14 &  $-$0.05 &  \nodata &  $-$0.74 &  $-$0.20 &    6 \\
HE~0454$-$4758	  &  $-$3.10 &  \nodata &  $+$0.29 &  $+$0.28 &  $+$0.32 &  $+$0.07 &  $-$0.11 &  $-$0.69 &  \nodata &    3 \\
CS~29502$-$092    &  $-$3.18 &  \nodata &  $+$0.42 &  $+$0.31 &  $+$0.19 &  $+$0.08 &  $-$0.63 &  $-$1.26 &  \nodata &    5 \\
CS~30325$-$094    &  $-$3.30 &  $+$0.09 &  $+$0.38 &  $+$0.38 &  $+$0.28 &  $+$0.04 &  \nodata &  $-$1.88 &  \nodata &    6 \\
HE~1337$+$0012	  &  $-$3.44 &  \nodata &  $+$0.55 &  $+$0.48 &  $+$0.51 &  $+$0.17 &  \nodata &  $+$0.07 &  \nodata &    3 \\
HE~1351$-$1049	  &  $-$3.45 &  \nodata &  $+$0.30 &  $+$0.32 &  $+$0.30 &  $+$0.30 &  \nodata &  $+$0.13 &  \nodata &    3 \\
CS~22172$-$002    &  $-$3.86 &  $-$0.35 &  $+$0.20 &  $+$0.37 &  $+$0.21 &  $-$0.15 &  $-$0.98 &  $-$1.17 &  \nodata &    6 \\
CD~$-$38~245      &  $-$4.20 &  $-$0.06 &  $+$0.20 &  $+$0.20 &  $+$0.28 &  $-$0.19 &  $-$0.45 &  $-$0.76 &  \nodata &    6 \\
\enddata					  				        
\tablerefs{
(1)~\citealt{stephens02};
(2)~\citealt{fulbright00};
(3)~\citealt{barklem05};
(4)~\citealt{ivans03};
(5)~\citealt{lai08};
(6)~\citealt{cayrel04}, \citealt{francois07};
(7)~\citealt{honda04b}
}
\end{deluxetable}

\clearpage

\begin{deluxetable}{lrrrrrrrrrrrrr} 
\tablecaption{Space Velocities and Orbital Parameters for the Globular 
Cluster Sample
\label{gctab}}
\tablewidth{0pt}
\tabletypesize{\scriptsize}
\tablehead{
\colhead{Cluster} &
\colhead{$\langle$[Fe/H]$\rangle$} &
\colhead{$D$} &
\colhead{$U$} &
\colhead{$V$} &
\colhead{$W$} &
\colhead{$V_{\phi}$} &
\colhead{$R_{\rm peri}$} &
\colhead{$R_{\rm apo}$} &
\colhead{$e$} &
\colhead{$|Z_{\rm max}|$} &
\colhead{$J_{X}$} &
\colhead{$J_{Y}$} &
\colhead{$J_{Z}$} \\
\colhead{} &
\colhead{} &
\colhead{(kpc)} &
\colhead{(\kmsec)} &
\colhead{(\kmsec)} &
\colhead{(\kmsec)} &
\colhead{(\kmsec)} &
\colhead{(kpc)} &
\colhead{(kpc)} &
\colhead{} &
\colhead{(kpc)} &
\colhead{(kpc \kmsec)} &
\colhead{(kpc \kmsec)} &
\colhead{(kpc \kmsec)}}
\startdata
         NGC~288 & $-$1.39 &   8.8 &   16 &  $-$27 &   53 &  $-$27 &  5.0 & 14.1 & 0.48 &  13.5 &  $-$238 & $-$1035 &  $-$456  \\
         NGC~362 & $-$1.33 &   8.5 &   21 &  $-$58 &  $-$80 &  $-$51 &  1.5 & 51.9 & 0.95 &  47.9 &    40 &  1478 & $-$1061  \\
  NGC~1904~(M79) & $-$1.42 &  12.9 &  120 &   32 &   71 &  103 &  6.1 & 18.8 & 0.51 &   6.4 &  $-$388 & $-$1423 &  1297  \\
        NGC~2298 & $-$1.90 &  10.7 &  $-$62 &   12 &   97 &  $-$27 &  3.8 & 16.4 & 0.62 &   3.2 &  $-$876 & $-$1046 &  $-$430  \\
  NGC~4590~(M68) & $-$2.34 &  10.2 &  188 &  254 &    5 &  301 &  2.0 & 33.0 & 0.89 &  26.0 & $-$1560 &  1022 &  6694  \\
   NGC~5272~(M3) & $-$1.50 &  10.4 &  $-$17 &  104 & $-$126 &  105 &  5.3 & 16.9 & 0.52 &  16.8 & $-$1237 &  2158 &  1948  \\
        NGC~5466 & $-$2.05 &  15.9 &  248 &   81 &  207 &   44 &  9.0 & 23.1 & 0.44 &  23.0 &  $-$610 &  $-$432 &   900  \\
        NGC~5897 & $-$1.84 &  12.4 &   30 &  $-$81 &  118 &  $-$77 &  7.4 & 12.7 & 0.26 &   7.6 &   136 & $-$3024 & $-$2110  \\
   NGC~5904~(M5) & $-$1.30 &   7.5 & $-$323 &   80 & $-$204 &   84 &  3.9 &  9.7 & 0.43 &   8.0 &  $-$493 &  2764 &  1865  \\
  NGC~6093~(M80) & $-$1.73 &  10.0 &  $-$12 &  $-$65 &  $-$81 &  $-$65 &  6.5 &  8.7 & 0.15 &   3.4 &   312 &  2099 & $-$1730  \\
   NGC~6121~(M4) & $-$1.18 &   2.2 &  $-$57 &   27 &   $-$8 &   26 &  0.1 &  1.0 & 0.88 &   0.7 &   $-$14 &   119 &   499  \\
  NGC~6218~(M12) & $-$1.36 &   4.9 &  $-$50 &  122 & $-$106 &  125 &  2.0 &  5.0 & 0.43 &   3.0 &  $-$396 &  2137 &  2646  \\
  NGC~6254~(M10) & $-$1.52 &   4.4 &  $-$84 &  133 &   97 &  137 &  1.6 &  4.1 & 0.43 &   2.3 &  $-$119 & $-$2170 &  2872  \\
        NGC~6362 & $-$1.04 &   7.6 &   81 &  108 &   37 &  121 &  2.9 &  9.6 & 0.54 &   3.0 &    97 & $-$1034 &  2805  \\
        NGC~6397 & $-$2.02 &   2.3 &   34 &  129 &  $-$99 &  130 &  0.1 &  1.5 & 0.93 &   0.7 &   144 &  1874 &  2491  \\
        NGC~6528 & $-$0.06 &   7.9 &   27 &  $-$26 & $-$227 &  $-$26 &  5.0 &  6.0 & 0.09 &   0.6 &   $-$61 &  5636 &  $-$653  \\
        NGC~6553 & $-$0.28 &   6.0 &    9 &  215 &   14 &  215 &  3.7 &  4.6 & 0.10 &   0.3 &    72 &  $-$323 &  4919  \\
  NGC~6656~(M22) & $-$1.49 &   3.2 &  152 &  195 & $-$118 &  191 &  0.8 &  1.7 & 0.38 &   0.5 &    19 &  2311 &  3844  \\
        NGC~6752 & $-$1.54 &   4.0 &   36 &  197 &   24 &  199 &  1.0 &  4.4 & 0.63 &   2.6 &   301 &  $-$548 &  4050  \\
  NGC~6809~(M55) & $-$1.88 &   5.3 & $-$202 &  115 & $-$181 &  121 &  2.9 &  4.8 & 0.25 &   2.6 &   115 &  4370 &  2648  \\
  NGC~6838~(M71) & $-$0.76 &   4.0 &  $-$77 &  162 &   $-$2 &  173 &  0.1 &  5.6 & 0.96 &   0.4 &    42 &    62 &  3364  \\
  NGC~7078~(M15) & $-$2.38 &  10.3 & $-$148 &    1 &  $-$58 &   56 &  1.4 & 14.0 & 0.81 &   5.1 &  $-$477 &  1908 &  1249  \\
  NGC~7099~(M30) & $-$2.31 &   8.0 &   68 &  $-$80 &   60 &  $-$87 &  3.1 & 11.8 & 0.58 &  10.1 &  $-$322 & $-$1715 & $-$1922  \\
           Pal~5 & $-$1.31 &  23.2 &   10 &  $-$42 &  $-$45 &  $-$42 & 19.0 & 21.9 & 0.07 &  16.7 &   692 &  1661 & $-$1396  \\
          Pal~12 & $-$0.75 &  19.1 & $-$225 & $-$109 &  $-$21 &  $-$55 & 11.0 & 21.9 & 0.33 &  17.5 & $-$1673 &  3763 & $-$1600  \\
\enddata
\end{deluxetable}

\end{document}